\documentclass[onecolumn,amsmath,amssymb,floatfix,showpacs,superscriptaddress]{revtex4}
\usepackage{graphicx,bm}

\begin{document}

\def\pdag{p\hspace{-1.5mm} / }
\def\tpdag{p\hspace{-1.3mm} / }
\def\qdag{q\hspace{-1.7mm} / }
\def\tqdag{q\hspace{-1.4mm} / }
\def\kdag{k\hspace{-1.9mm} / }
\def\tkdag{k\hspace{-1.5mm} / }
\def\edag{\varepsilon\hspace{-1.7mm} / }

\title{Unitary model for the $\bm{\gamma p \to \gamma \pi^0 p}$ reaction and \\
              the magnetic dipole moment of the $\Delta^+(1232)$}

\author{Wen-Tai Chiang}
\affiliation{Department of Physics and National Center for Theoretical Sciences at Taipei, National Taiwan University, Taipei 10617, Taiwan}
\affiliation{Institute of Physics, Academia Sinica, Taipei 11529, Taiwan}

\author{M. Vanderhaeghen}
\affiliation{Jefferson Laboratory, Newport News, VA 23606, USA}
\affiliation{Department of Physics, College of William and Mary, Williamsburg, VA 23187, USA}

\author{Shin Nan Yang}
\affiliation{Department of Physics and National Center for Theoretical Sciences at Taipei, National Taiwan University, Taipei 10617, Taiwan}

\author{D. Drechsel}
\affiliation{Institut f\"ur Kernphysik, Universit\"at Mainz, 55099 Mainz, Germany}

\date{\today}

\begin{abstract}
Radiative pion photoproduction in the $\Delta(1232)$ resonance region is
studied with the aim to access the $\Delta^+(1232)$ magnetic dipole moment. We
present a unitary model of the $\gamma p \to \gamma \pi N$ ($\pi N = \pi^0
p,\;\pi^+ n$) reactions, where the $\pi N$ rescattering is included in an
on-shell approximation. In this model, the low energy theorem which couples the
$\gamma p \to \gamma \pi N$ process in the limit of a soft final photon to the
$\gamma p \to \pi N$ process is exactly satisfied. We study the sensitivity of
the $\gamma p \to \gamma \pi^0 p$ process at higher values of the final photon
energy to the $\Delta^+(1232)$ magnetic dipole moment. We compare our results
with existing data and give predictions for forthcoming measurements of angular
and energy distributions. It is found that the photon asymmetry and a helicity
cross section are particularly sensitive to the $\Delta^+$ magnetic dipole
moment.
\end{abstract}
\pacs{12.39.Jh, 13.60.Fz, 14.20.Gk}

\maketitle

\section{Introduction}
\label{sec:intro}

The $\Delta(1232)$ is the first and most prominent excited state of the nucleon
and the only well isolated nucleon resonance. Its  properties provide an
important test for theoretical descriptions in the non-perturbative domain of
QCD.
There are two kinds of electromagnetic properties of the $\Delta$. The first
one involves the $N\rightarrow\Delta$ transition, described by the magnetic
dipole ($\mu_{N\Delta}$) and electric quadrupole ($Q_{N\Delta}$) transition
moments to be determined from pion electromagnetic
production~\cite{Blanpied01,Yang03}. The other properties involve the $\Delta$
itself, the magnetic dipole moment $\mu_{\Delta}$, the electric quadrupole
moment $Q_{\Delta}$, and the magnetic octupole moment of the resonance. They
are difficult to measure because of the short life time of the $\Delta$.
\newline
\indent
In particular, the magnetic dipole moment (MDM) of the $\Delta(1232)$ is of
considerable theoretical interest. In symmetric SU(6)  quark models, the
nucleon and the $\Delta$ resonance are degenerate and their magnetic moments
are related through $\mu_\Delta = e_\Delta \, \mu_p$, where $e_\Delta$ is the
electric charge of the $\Delta$, and $\mu_p$ the proton magnetic moment.
However, different theoretical models predict considerable deviations from this
SU(6) value \cite{Ali00b}.
The $\Delta(1232)$ MDM has also been investigated on the lattice at rather
large quark masses \cite{Leinweber92}, and very recently the chiral
extrapolation of the $\Delta(1232)$ MDM, including the next-to-leading
non-analytic variation with the quark mass, was also studied
\cite{Cloet02}. At present, there still is a considerably large gap
in quark mass to bridge between the state-of-the-art
lattice QCD calculations and the chiral limit. Therefore, it would be
extremely helpful to know the resonance MDM for the physical quark mass
values, through experiment. Unfortunately, the experimental information on
the MDMs beyond the ground state baryon octet is very scarce. With the
notable exception of the $\Omega^-$ baryon, these higher nucleon resonances
decay strongly, and thus have too short lifetimes to measure their MDMs in
the conventional way through spin precession measurements.
\newline
\indent
The magnetic moment of
the $\Delta^{++}(1232)$ has been  measured by the reaction
$\pi^+ p \to \gamma \pi^+ p$ \cite{Nef78,Bos91}.
As a result of these measurements,
and using different theoretical
analyses, the PDG \cite{PDG02} quotes the range~:
$\mu_{\Delta^{++}} = 3.7 - 7.5 \, \mu_N$ (where $\mu_N$ is the
nuclear magneton), while
SU(6) symmetry results in the value $\mu_{\Delta^{++}} = 5.58 \, \mu_N$.
The large uncertainty in the extraction of the experimental value
is due to large non-resonant contributions to the
$\pi^+ p \to \gamma \pi^+ p$ reaction because of bremsstrahlung from the
charged pion ($\pi^+$) and proton (p).
\newline
\indent
For the $\Delta^+(1232)$, it has been proposed \cite{Dre83} to
determine its magnetic moment through measurement of the $\gamma p \to \gamma
\pi^0 p$ reaction. Due to the small cross sections for this reaction, which is
proportional to $\alpha_{em}^2 = 1/(137)^2$, a first measurement has only
recently been reported by the A2/TAPS collaboration at MAMI \cite{Kot02}.
At present, dedicated experiments are being performed with much higher count
rates by using $4 \pi$ detectors, such as the
Crystal Ball detector at MAMI \cite{CB}. The analysis of this next
generation of dedicated experiments requires a substantial theoretical effort
aimed at minimizing the model dependencies in the extraction of the
$\Delta^+$ MDM from the measurement of the $\gamma p \to \gamma \pi^0 p$
observables.
\newline
\indent
First estimates for the reaction $\gamma p \to \gamma \pi^0 p$,
including only the $\Delta$-resonant mechanism, were performed in
Refs.~\cite{Mac99,DVGS}. An improved calculation which contains both the
$\Delta$-resonant mechanism and a background of nonresonant contributions has
subsequently been carried out in Ref.~\cite{DV01}.
The starting point of the model is an
effective Lagrangian description of the $\gamma p \to  \pi^0 p$ process.
Then an additional photon is coupled
in a gauge invariant way to describe the
$\gamma p \to \gamma \pi^0 p$ reaction. The result is a tree level calculation
with part of the final state interaction effects taken into account by the
finite width of the $\Delta$. This model was used in the analysis of the
pioneering measurement of the $\gamma p \to \gamma \pi^0 p$ cross sections
and an initial value of
$\mu_{\Delta^+} = \left[ 2.7 {{+1.0} \atop {-1.3}}
(\mathrm{stat.}) \pm 1.5 (\mathrm{syst.}) \pm 3 (\mathrm{theor.}) \right]
\, \mu_N$
has been extracted in Ref.~\cite{Kot02}.
\newline
\indent
Although the tree level model of Ref.~\cite{DV01}
gives a qualitatively good description of the data of Ref.~\cite{Kot02},
a detailed quantitative comparison requires the
inclusion of rescattering effects. Such rescattering effects were
found to be important in the case of pion photoproduction
(see e.g.~\cite{Yang91}). Since an accurate theoretical description of the
reaction $\gamma p \to \gamma \pi^0 p$ is essential for extracting a precise
value for $\mu_{\Delta^+}$, it is imperative to obtain an estimate of the
effects of the final-state interaction to the best of our capability. It is
therefore the aim of our present work to describe the radiative pion
photoproduction by a properly unitarized theory.
\newline
\indent We start in Sec.~II  by specifying the kinematics and cross section of
the reaction $\gamma p \to \gamma \pi^0 p$. In Sec.~III, we present a unitary
model for the $\gamma p \to  \pi^0 p$ process with a transition potential that
is derived from an effective Lagrangian with Born terms and vector mesons
exchange in addition to the $\Delta$-excitation mechanism. Our model is very
similar to MAID~\cite{MAID} in that only the on-shell rescattering effects are
included in the nonresonant multipoles.  We further show a selection of our
fits to various experimental data, which have been used to fix all the
parameters of the strong interaction. Our model for the $\gamma \, p \, \to \,
\gamma \, \pi^0 \, p$ reaction in the $\Delta(1232)$-resonance region
is described in Sec.~IV. The transition potential for this reaction is given by
all the tree diagrams that can be obtained from the effective Lagrangian
previously adopted for the $\gamma p \to \pi^0 p$ process, with the addition of
the anomaly terms generated from the $\pi^0 \to \gamma + \gamma$ vertex. We
then proceed to estimate the final-state interaction effects by including the
on-shell rescattering between pion and nucleon. In Sec.~V we compare our
results for the $\gamma \, p \, \to \, \gamma \, \pi^0 \, p$ reaction
with the existing data. We further present our predictions for several angular
and energy distributions as well as polarization observables that are expected
to be measured in forthcoming experiments. In each case we demonstrate the
sensitivity with respect to the $\Delta^+(1232)$ magnetic dipole moment. We
close by summarizing our findings in Sec.~VI.

\section{KINEMATICS AND CROSS SECTION FOR THE  $\gamma p \to \gamma \pi^0 p$
REACTION}
\label{sec2:form}

In the $\gamma p \to \gamma \pi N$ process, a photon ($k$, $\lambda$) hits a
proton target ($p$, $s_N$), leading to a final state with a photon ($k'$,
$\lambda'$), a pion ($q$), and a proton or neutron ($p'$, $s'_N$). Here $k$,
$k'$, $p$, $p'$, and $q$ are the four-momenta of the respective particles,
$\lambda$ and $\lambda'$ denote the photon helicities, and $s_N$ and $s'_N$ are
the nucleon spin projections.
\newline
\indent
Our results for the experimental observables will be expressed in the
center-of-mass ({\it c.m.})
frame of the initial $\gamma p$ system with  total {\it c.m.}
energy squared given by the usual Mandelstam invariant $s = (k + p)^2$. The
kinematics of the $\gamma p \to \gamma \pi N$ reaction can be described by 5
variables.  First, we choose the energies $E_\gamma$ and $E^{\prime}_\gamma$ of
the initial and outgoing photon, respectively. The other three variables  are
the polar ( $\theta_\gamma$ ) and azimuthal ( $\phi_\gamma$ )
angles of the final photon,
and $\theta_\pi$, the polar angle of the pion. These angles are
defined with regard to an x-z plane which contains the initial particles and
the final pion, with the photon momentum $\bm{k}$ pointing in the $z$-direction
and $\phi_\pi \equiv 0$.
\newline
\indent
The unpolarized five-fold differential cross section for the $\gamma p \to
\gamma \pi N$ reaction, differential with respect to the outgoing photon energy
and angles as well as the pion angles in the {\it c.m.} system, takes the form
\begin{eqnarray}
\label{eq:crosssection}
\left( \frac{d\sigma}{dE^{\prime}_\gamma\,
                d\Omega'_\gamma\,d\Omega_\pi} \right)^{c.m.}\;
&=&\;
 \frac{1}{(2 \pi)^5}\, {\frac{1}{32 \sqrt{s}}}\,
 \frac{E^{\prime}_\gamma}{E_\gamma} \,
 \frac{|\,\bm{q}\,|^2}{|\,\bm{q}\,| (E'_N+\omega_q) +
 E^{\prime}_\gamma \omega_q \cos \theta_{\gamma' \pi} } \, \nonumber\\
&\times&
 \left(\, \frac{1}{4}\sum_{\lambda}\sum_{s_N}\sum_{\lambda'}\sum_{s'_N}
 | \, \varepsilon_\mu(k, \lambda) \, \varepsilon_\nu^*(k',\lambda') \,
 {\mathcal M}^{\nu \mu} \, |^2 \, \right) \, .
\end{eqnarray}
Unless otherwise specified, $E_\gamma$ and $E^{\prime}_\gamma$ refer to
the initial and final photon energies in the {\it c.m.} system.
Furthermore, $\omega_q$ and $\bm{q}$ denote
the energy and momentum of the pion, $E'_N$ the
final nucleon energy, $\theta_{\gamma' \pi}$ the {\it c.m.} angle between the
outgoing photon and the pion, and $\varepsilon_\mu(k,\lambda)$ and
$\varepsilon_\nu^*(k',\lambda')$ are the polarization vectors
of the incoming and outgoing photons, respectively.
Furthermore, ${\mathcal M}^{\nu \mu}$ is a
tensor for the $\gamma p \to \gamma \pi N$ process, which will be discussed
in Sec.~IV.
\newline
\indent
We will also show results for partially integrated cross sections of the
$\gamma p\to\gamma\pi^0 p$ reaction, e.g., the cross section
$d\sigma/dE^{\prime}_\gamma$ differential with respect to the outgoing
 photon {\it c.m.} energy, or the cross section
$d\sigma/dE^{\prime}_\gamma d\Omega^{c.m.}_\pi$
differential with respect to the outgoing photon {\it c.m.} energy
and the pion {\it c.m.} solid angle. These cross
sections are obtained by integrating the fully differential cross section of
Eq.~(\ref{eq:crosssection}) over the appropriate part of the phase space.

\section{Unitary model for the $\gamma p \to \pi N$ reaction}
\label{sec:pi}

\begin{figure}[tbp]
\begin{center}
\includegraphics[width=0.75\columnwidth]{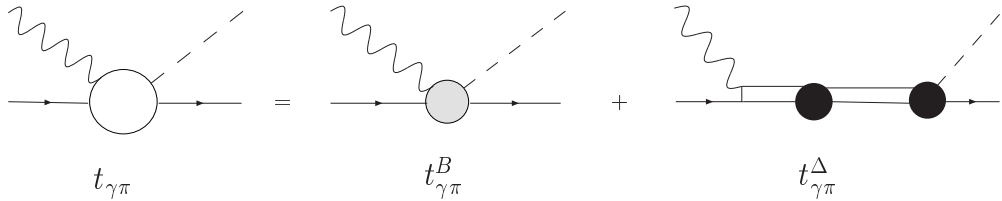}
\end{center}
\caption{ Graphical representation of the pion photoproduction $T$-matrix.
\label{fig1}}
\end{figure}

\begin{figure}
\centering
\includegraphics[width=0.3\columnwidth]{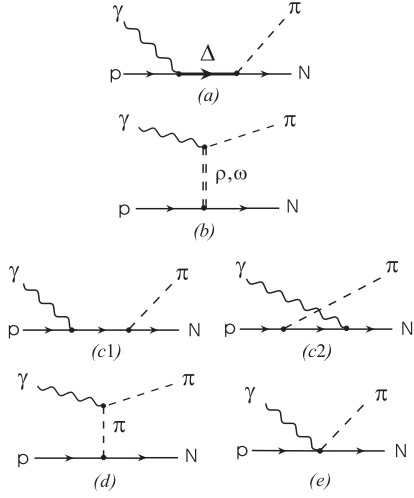}
\caption{\label{fig:gap_piN_diag}Diagrams for the $\gamma p \rightarrow \pi N$
reaction in the $\Delta(1232)$ region: $\Delta$ resonance excitation (a),
vector meson exchange (b), nucleon pole terms (c1-c2), pion pole term (d), and
Kroll-Rudermann term (e). }
\end{figure}

In the dynamical approach to pion photo- and electroproduction~\cite{Yang85},
where unitarity is built in by explicit inclusion of the final state $\pi N$
interaction, the $T$-matrix is expressed as
\begin{eqnarray} \label{eq:tmatrix}
t_{\gamma\pi} = v_{\gamma\pi} + v_{\gamma\pi} g_0 t_{\pi N},
\end{eqnarray}
where $v_{\gamma\pi}$ is a transition operator for  the reaction $\gamma  N \to
\pi N$, and $t_{\pi N}$ and $g_0$ denote the $\pi N$ scattering matrix and the
free propagator, respectively.
If the on-shell or $K$-matrix approximation is made, that is the intermediate
particles (pions and nucleons) are restricted to be on the mass shell, the
magnitudes of the on-shell momenta for the intermediate particles depend only
on the total {\it c.m.} energy $W_{\pi N}$
of the $\gamma N \to \pi N$ process. We
therefore obtain the following expression for
 the physical amplitude in the {\it c.m.} frame:
\begin{eqnarray} \label{eq:onBSeq}
   t_{\gamma \pi}(\bm{q},\bm{k}; W_{\pi N})
   = v_{\gamma \pi }(\bm{q},\bm{k})
    - \frac{i}{32\pi^2} \frac{\,|\bm{q}|\,}{W_{\pi N}}
      \sum_{s'_N} \int d\Omega_{q'}\,
      T_{\pi N}(\bm{q},-\bm{q};\bm{q'},-\bm{q'})\,
      v_{\gamma \pi }(\bm{q'},\bm{k}) \,,
\end{eqnarray}
where we sum over the final nucleon spins $s'_N$ and the relevant $\pi N$
channels. For example, in the case of $ \gamma p \to \pi^0 p$ we need to
include both $\pi^0 p$ and $\pi^+ n$ intermediate states. The Lorentz invariant
$T$-matrix is given by \cite{Hamilton60}
\begin{eqnarray} \label{eq:tpin}
  T_{\pi N}(\bm{q'},\bm{p'};\bm{q},\bm{p})
  = \bar{u}(p',s'_N) \left[ \,A + \tfrac{1}{2}(\qdag+\qdag')\,B\, \right]
  u(p,s_N) \,,
\end{eqnarray}
where $A$ and $B$ are scalar functions of the invariants $s=W_{\pi N}^2$ and
$t$, the square of the four-momentum transfer.
We use the covariant normalization
$\bar u u = 2 M_N$ ($M_N$ denotes the nucleon mass)
for the Dirac spinors, and construct the functions A and B
from the SAID partial wave amplitudes $f_{\ell\pm}(W_{\pi N})$.
\newline
\indent
For pion photoproduction in the $\Delta(1232)$ resonance region,
the transition potential $v_{\gamma \pi}$ consists of two terms
\begin{equation} \label{eq:vBD}
  v_{\gamma \pi} = v_{\gamma \pi}^B  + v_{\gamma \pi}^\Delta \,,
\end{equation}
where $v_{\gamma  \pi}^\Delta$ corresponds to the resonance contribution
$\gamma N \to \Delta \to \pi N$ and $v_{\gamma \pi}^B$ describes the background
to be derived from an effective Lagrangian.
 The resulting $T$-matrix can be decomposed into two
terms, as shown in Fig.~\ref{fig1}~\cite{KY99}:
\begin{eqnarray}\label{eq:KY}
t_{\gamma\pi} =t_{\gamma\pi}^B + t_{\gamma\pi}^{\Delta}\,,
\label{eq:ddecomp}
\end{eqnarray}
where
\begin{eqnarray}\label{eq:tdecomp}
t_{\gamma\pi}^B(W_{\pi N})&=&v_{\gamma\pi}^B+v_{\gamma\pi}^B\,g_0(W_{\pi N})
\,t_{\pi N}(W_{\pi N}), \\
t_{\gamma\pi}^\Delta(W_{\pi N})&=&v_{\gamma\pi}^\Delta+v_{\gamma\pi}^\Delta\,
g_0(W_{\pi N}) \,t_{\pi N}(W_{\pi N})\,. \label{eq:decomp}
\end{eqnarray}
The solid blobs in Fig.~\ref{fig1} indicate that both the intermediate $\Delta$
states and the $\pi N\Delta$ vertices are both dressed~\cite{Hsiao98,Tanabe85}.
%
\newline
\indent
Applying Eq.~(\ref{eq:onBSeq}) to the background contribution
$t_{\gamma \pi}^B$, we obtain
\begin{eqnarray} \label{eq:onBSeq1}
  t_{\gamma \pi}^B(\bm{q},\bm{k}; W_{\pi N})
  = v_{\gamma\pi}^B(\bm{q},\bm{k})
    - \frac{i}{32 \, \pi^2} \frac{\,|\bm{q}|\,}{W_{\pi N}}
      \sum_{s'_N} \int d\Omega_{q'}\,
      T_{\pi N}(\bm{q},-\bm{q};\bm{q'},-\bm{q'})\,
      v_{\gamma\pi }^B(\bm{q'},\bm{k}) \, .
\end{eqnarray}
\newline
\indent
Due to the on-shell approximation, the multipole amplitudes
$t^{B,\,\alpha}_{\gamma \pi}$, for the partial wave $\alpha$,
in Eq.~(\ref{eq:onBSeq1}), take the form
\begin{eqnarray} \label{eq:cose}
  t^{B,\,\alpha}_{\gamma \pi}
  = v^{B,\,\alpha}_{\gamma\pi}\, \cos\,\delta_\alpha\,
  e^{i \delta_\alpha} \,,
\end{eqnarray}
where $\delta_\alpha$ is the phase shift for $\pi N$ scattering in the
respective partial wave $\alpha$.
\newline
\indent
The resonance structure $t_{\gamma\pi}^{\Delta}(E)$, as depicted in Fig. 1, is
approximated by
\begin{equation} \label{eq:Duni1}
  t_{\gamma\pi}^\Delta = v_{\gamma\pi}^\Delta
(M_\Delta\to M_\Delta- \frac{i}{2} \Gamma_\Delta)  e^{i \phi(W_{\pi N})} \,,
\end{equation}
with the phase $\phi(W_{\pi N})$ adjusted such that the $\Delta$ multipole
amplitudes ($M_{1+}^\Delta,\ E_{1+}^\Delta$) carry the phase of the $\pi N$
scattering phase $\delta_{33}(W_{\pi N})$ to ensure that the Fermi-Watson
theorem is fulfilled. We further adopt the ``complex mass scheme'' by
substituting $M_\Delta\to M_\Delta-\frac{i}{2} \Gamma_\Delta$ with an energy
independent width $\Gamma_\Delta$, as was suggested by Refs.
\cite{DV01,ElAmiri:1992xa,LopezCastro:2000ep} in order to maintain the gauge
invariance of the $\Delta$ contribution to the $\gamma N\to \gamma\pi N$
reaction. Since the background contribution of Eq.~(\ref{eq:cose}) satisfies
the Fermi-Watson theorem separately, also the total multipole amplitude will
carry the proper $\pi N$ phase.
\newline
\indent
In this work we focus on the energy region of the $\Delta(1232)$ resonance, and
neglect the contribution from the higher resonances. Although this is an
excellent approximation close to the $\Delta(1232)$ region, it is inevitable
that deviations will occur if we move further away from the $\Delta(1232)$
peak. The $P_{11}(1440)$ resonance is the nearest nucleon resonance. Due to its
large decay width, it is the most likely candidate to contribute to pion
photoproduction on the high-energy tail of the $\Delta(1232)$. Since the
$P_{11}(1440)$ contributes to the $M_{1-}^{(1/2)}$ multipole, we indeed find
deviations for this multipole between our calculation and the data if we
approach the $P_{11}(1440)$ energy region.
\newline
\indent
As has been outlined in Ref.~\cite{MAID}, we describe the non-resonant
transition operator $v^B_{\gamma\pi}$ by the tree diagrams of
Figs.~\ref{fig:gap_piN_diag}(b-e), as prescribed by an effective Lagrangian.
The electromagnetic $\gamma NN$ and $\gamma\pi\pi$ vertices are well known,
\begin{eqnarray} \label{eq:Lagem}
\mathcal{L}_{\gamma NN} &=&
  - e\,\bar\psi_N \left[ \,\hat{e}_N \gamma_{\mu} A^{\mu}
  - \frac{\kappa_N}{2 M_N}\, \sigma_{\mu\nu}\, \partial^{\nu} A^{\mu}
  \right] \psi_N \,,
   \nonumber\\ [0.5ex]
\mathcal{L}_{\gamma\pi\pi} &=&
  e \left[ (\partial_\mu\bm{\pi})^\dag \times \bm{\pi} \right]_3 A^\mu \,,
\end{eqnarray}
where $A^{\mu}$ is the electromagnetic vector potential, and $\psi_N $ and
$\bm{\pi} $ are the nucleon and pion field operators, respectively.
Furthermore, $\kappa_N$ is the nucleon anomalous magnetic moment
($\kappa_p=1.79$, $\kappa_n=-1.91$).
%
\newline
\indent
In the low energy regime addressed in this work, we
use the $\pi NN$ interaction Lagrangian with
pseudovector coupling (PV)~:
\begin{eqnarray} \label{eq:LagPV}
\mathcal{L}_{\pi N N}^{PV} &=&
  \frac{f_{\pi NN}}{m_\pi}\,\bar{\psi}_N\,
  \gamma_\mu\gamma_5\,\bm{\tau}\,\psi_N \cdot\partial^\mu\bm{\pi} \, ,
\end{eqnarray}
which is consistent with the leading order of chiral perturbation theory.
In Eq.~(\ref{eq:LagPV}), $\bm{\tau}$ are the Pauli (isospin) matrices,
and the coupling constant is taken as $f_{\pi NN}^2 / 4 \pi =0.081$.
At higher energies, an improvement of the non-resonant multipoles can
be obtained by a mixed pseudoscalar (PS) - pseudovector (PV) $\pi NN$
coupling as, e.g., in the MAID analysis~\cite{MAID}.
In the $\Delta(1232)$ resonance region considered in this work, we
prefer to stay consistent with the leading order
chiral perturbation theory and will use the PV coupling in the description of
both $\gamma p \to \pi N$ and $\gamma p \to \gamma \pi N$ reaction.
\newline
\indent
The relevant effective Lagrangians for the vector meson ($\rho$ and $\omega$)
exchanges are shown in Fig.~\ref{fig:gap_piN_diag}(b) and given by
\begin{eqnarray} \label{eq:LagV}
\mathcal{L}_{V\pi\gamma} &=&
  \frac{e g_{V\pi\gamma}}{m_{\pi}}\,
  \varepsilon_{\mu \nu \rho \sigma}\,(\partial^{\mu}A^{\nu})\,
  \pi_i\, \partial^{\rho}(\omega^{\sigma} \delta_{i3} + \rho_i^{\sigma}) \,,
  \nonumber\\ [0.5ex]
\mathcal{L}_{VNN} &=&
  g_{VNN} {\bar{\psi}_N} \left( \gamma_{\mu} V^{\mu} -
  \frac{\kappa_V}{2 M_N} \sigma_{\mu \nu} \partial^{\nu} V^{\mu}
  \right) \psi_N \,,
\end{eqnarray}
where $V$ denotes the $\rho$ and $\omega$ vector meson fields.
The photon couplings
$g_{\rho\pi\gamma}$ and $g_{\omega\pi\gamma}$ can be obtained from the
radiative decays $\rho \rightarrow \gamma \pi$ and $\omega \rightarrow \gamma
\pi$, which leads to the values $g_{\rho^+\pi\gamma}=0.103$,
$g_{\rho^0\pi\gamma}=0.131$, and $g_{\omega\pi\gamma}=0.314$. For the hadronic
couplings $g_{VNN}$ and $\kappa_{VNN}$, we use the values: $g_{\rho NN} =
2.63$, $\kappa_\rho = 6.1$, $g_\omega = 20$, and $\kappa_\omega = 0$. With
these effective Lagrangians, it is straightforward to derive the amplitude
shown in Fig.~\ref{fig:gap_piN_diag}(b).
%
\begin{figure}[h]
\includegraphics[width=0.45\columnwidth]{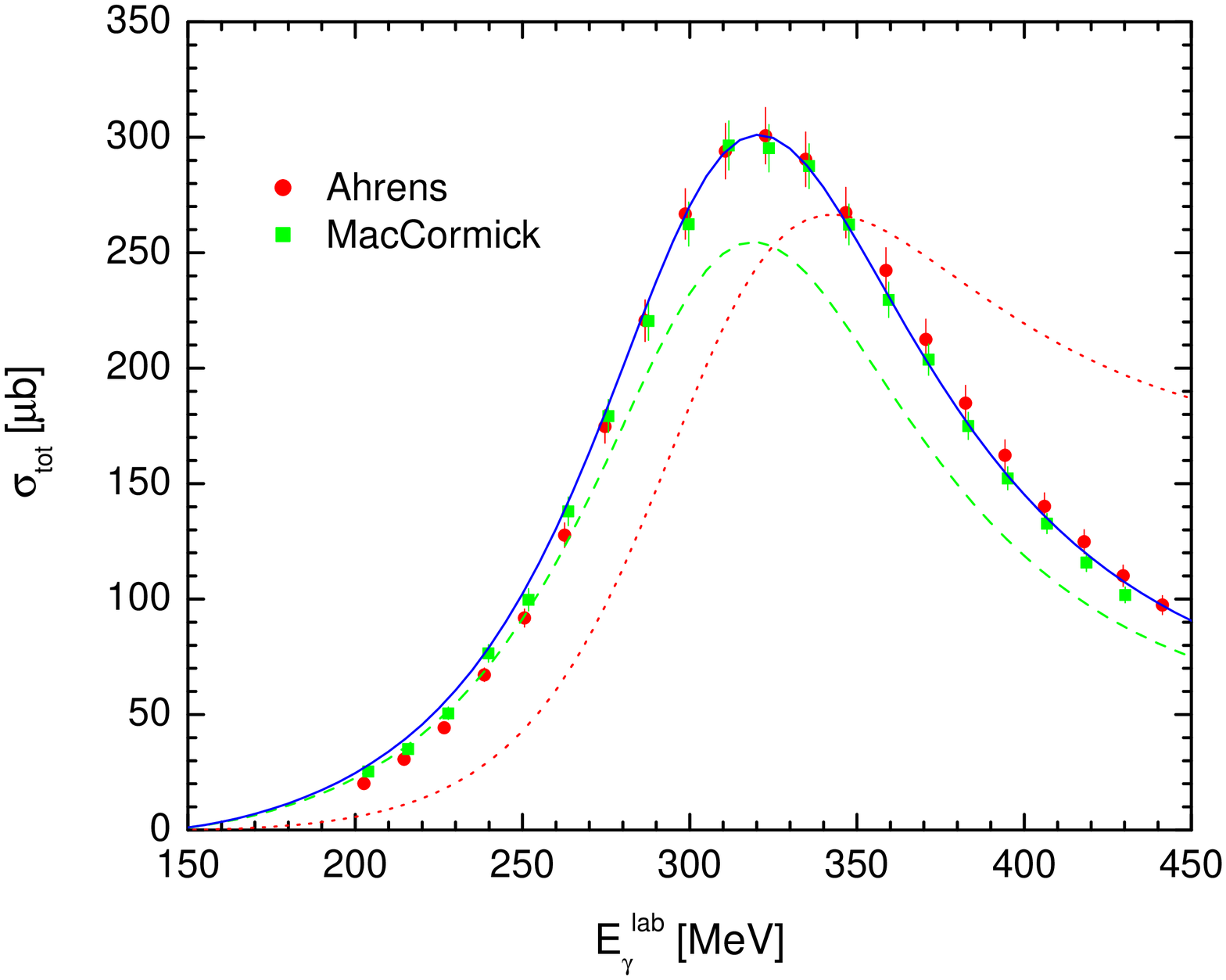}
\includegraphics[width=0.45\columnwidth]{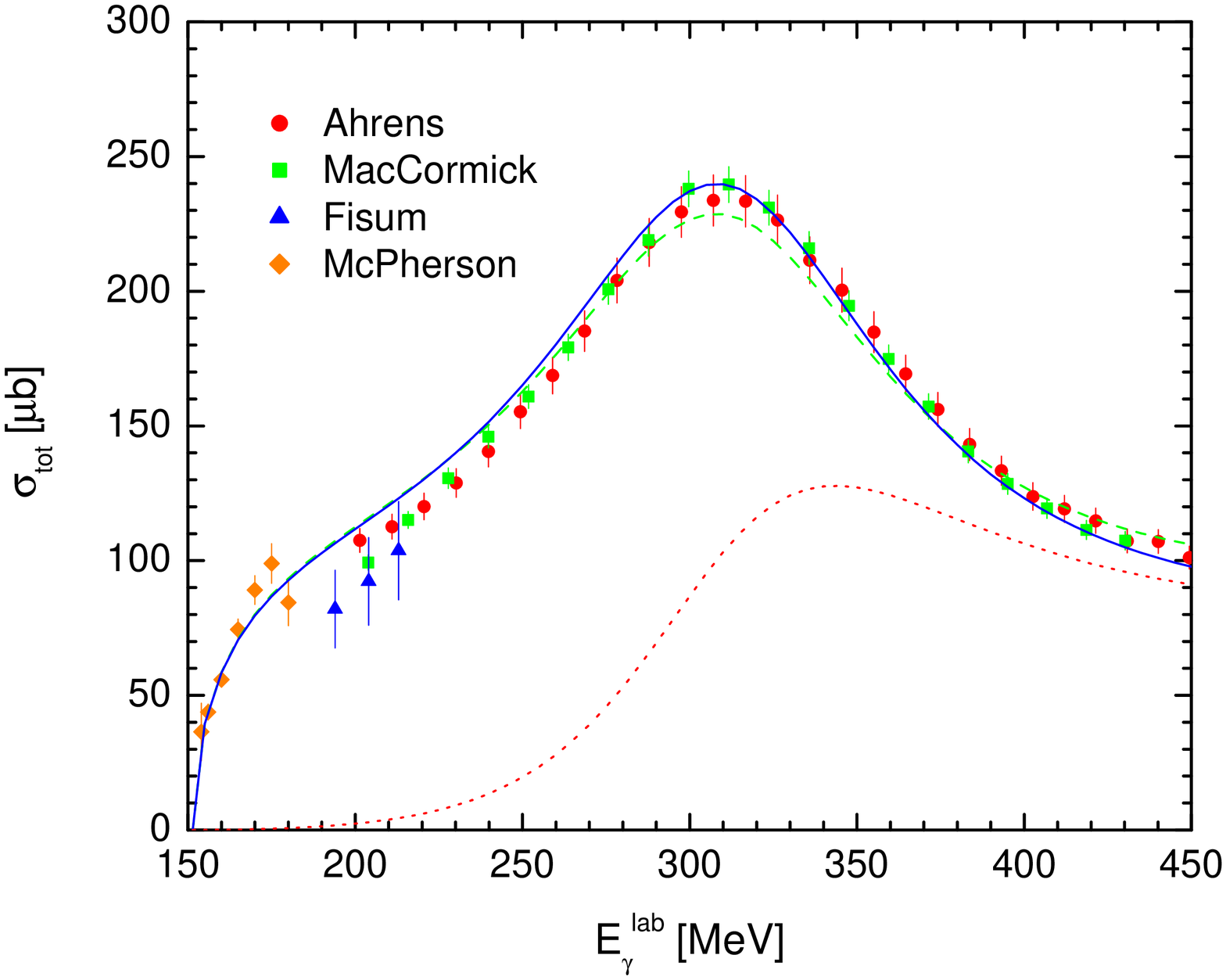}
\caption{\label{fig:gap_piop_tot}Total cross section for
$\gamma p \rightarrow \pi^0 p$ (left panel)
and $\gamma p \rightarrow \pi^+ n$ (right panel).
The solid curve is the full result of the unitary
model, the dashed curve indicates the result of the tree-level calculation, and
the dotted curve shows the unitarized $\Delta(1232)$ contribution.
The data for $\gamma p \rightarrow \pi^0 p$ are
from MacCormick~\cite{MacCormick:1996jz} and Ahrens~\cite{Ahrens:2000}.
The data for $\gamma p \rightarrow \pi^+ n$ are from
McPherson~\cite{McPherson:1964}, Fissum~\cite{Fissum:1996},
MacCormick~\cite{MacCormick:1996jz}, and Ahrens~\cite{Ahrens:2000}.}
\end{figure}
%
%
\begin{figure}[h]
\includegraphics[width=0.35\columnwidth]{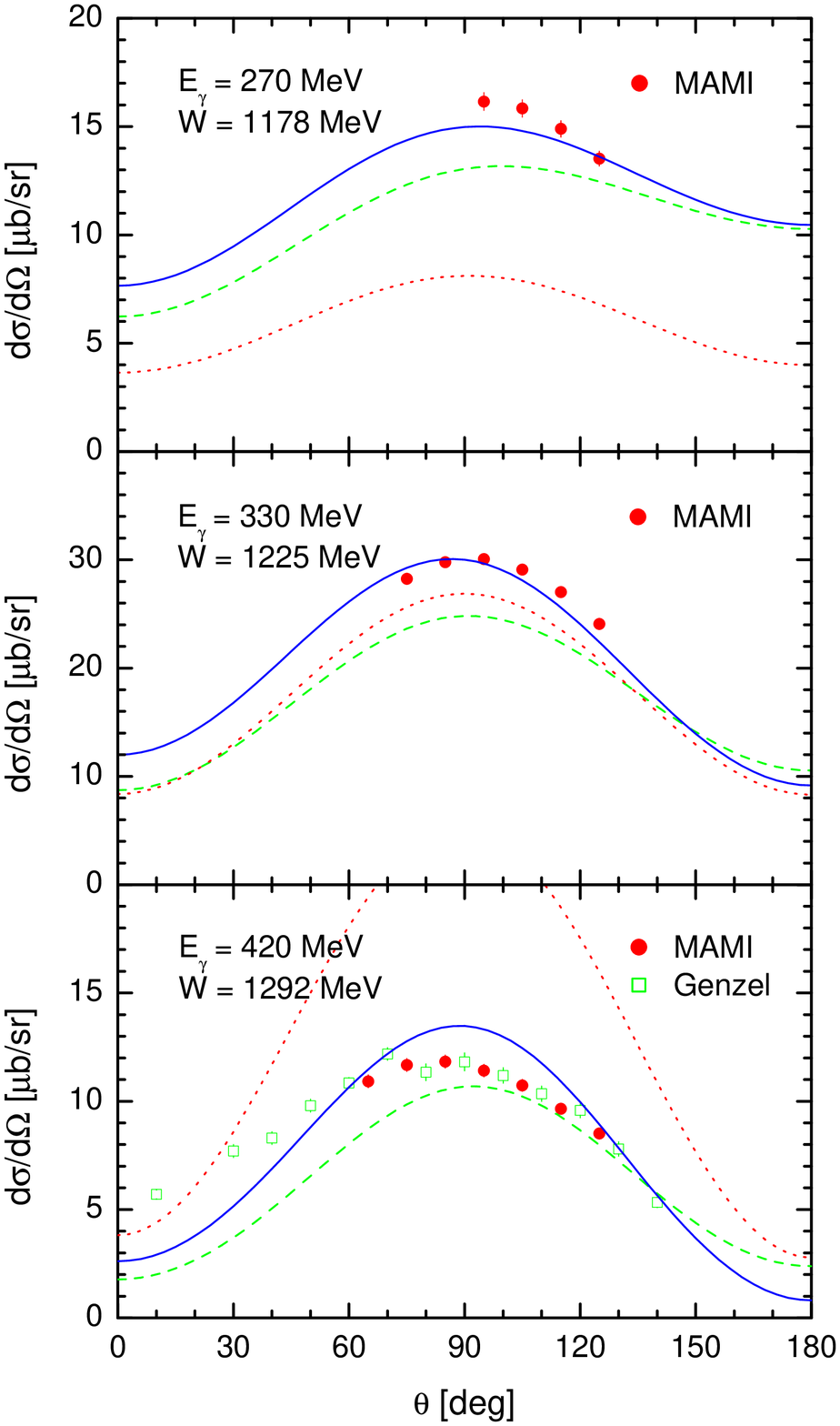}
\includegraphics[width=0.35\columnwidth]{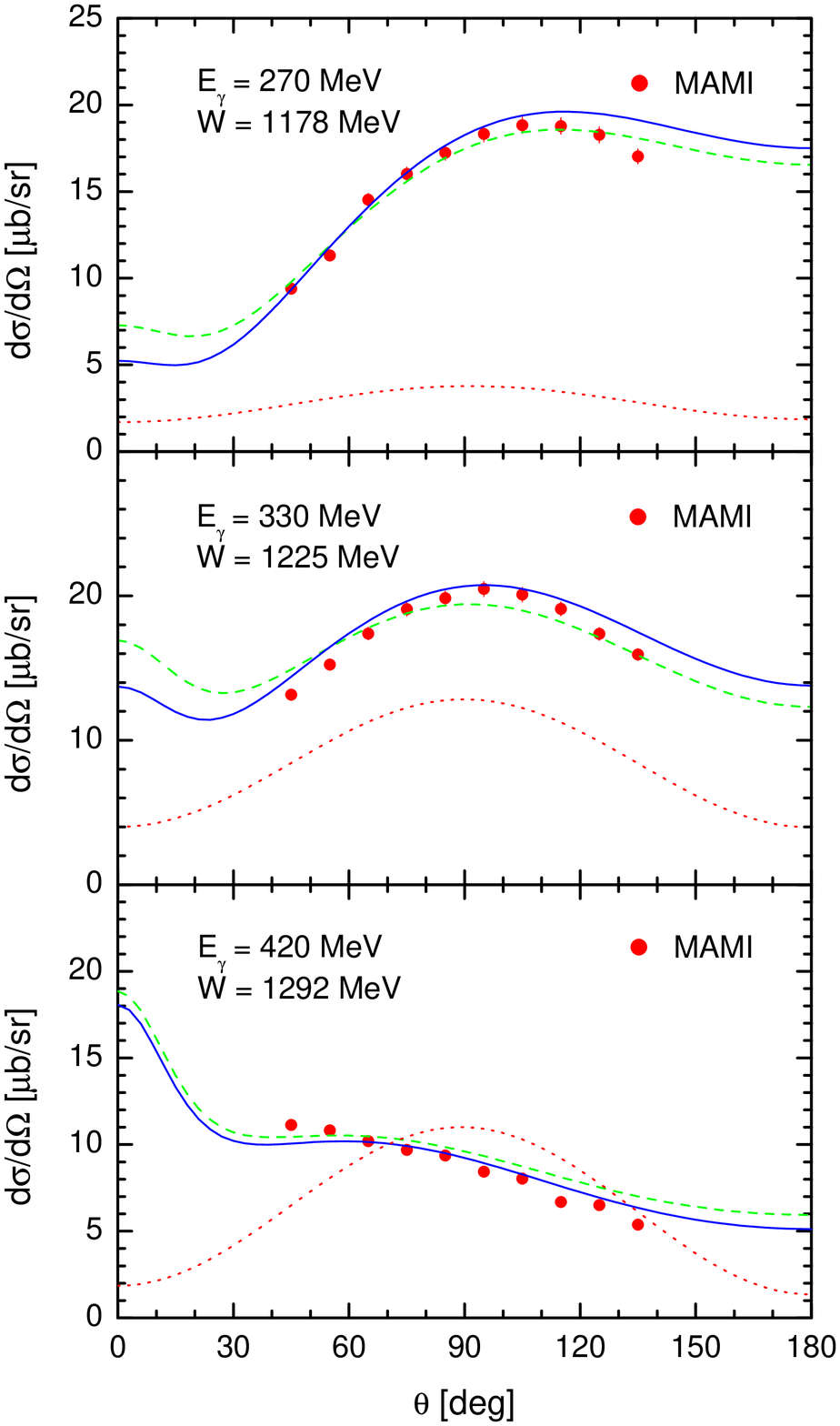}
\caption{\label{fig:gap_piop_diff}
Differential cross section for
the $\gamma p \rightarrow \pi^0 p$ reaction (left panel)
and $\gamma p \rightarrow \pi^+ n$ reaction (right panel)
at different photon {\it lab} energies $E_\gamma$
as function of the {\it c.m.} angle $\theta$.
The data are from MAMI~\cite{Beck:1999ge} and
Bonn~\cite{Genzel:1974}. See Fig.~\ref{fig:gap_piop_tot} for further notation.
}
\end{figure}
%
\newline
\indent
To calculate the $\Delta(1232)$ resonance contribution to
$v^\Delta_{\gamma\pi}$ of Fig.~\ref{fig:gap_piN_diag}(a), we use
the following form of the Rarita-Schwinger propagator~\cite{ElAmiri:1992xa}~:
\begin{eqnarray}
\label{eq:redprop}
  \tilde G_{\alpha \beta}(p_\Delta)
  &=& \frac{\pdag_\Delta + M_\Delta}{p_\Delta^2 - M_\Delta^2}\,
  \Bigg \{ - g_{\alpha \beta} \,+\, \frac{1}{3}\,\gamma_\alpha \gamma_\beta
  \,+\, \frac{1}{3 M_\Delta}\,\left(\gamma_\alpha (p_\Delta)_\beta
  \,-\, \gamma_\beta (p_\Delta)_\alpha \right)
  \,+\, \frac{2}{3 M_\Delta^2} \, (p_\Delta)_\alpha \, (p_\Delta)_\beta
  \,\Bigg\}
  \nonumber\\
  & & -\, \frac{2}{3 M_\Delta^2} \, \Bigg \{ \gamma_\alpha
  (p_\Delta)_\beta \,+\, \gamma_\beta (p_\Delta)_\alpha \,-\, \gamma_\alpha
  \, (\pdag_\Delta - M_\Delta) \, \gamma_\beta \Bigg \} \,,
\end{eqnarray}
where $p_\Delta$ is the four-momentum and $M_\Delta$ the mass of the
$\Delta(1232)$. The interaction Lagrangians for the vertices $\pi N\Delta$ and
$\gamma N\Delta$ are
\begin{eqnarray}
 \mathcal{L}_{\pi N\Delta} &=&
  \frac{f_{\pi N\Delta}}{m_\pi}\, \bar{\psi}_\Delta^\mu
  \,\bm{T}^\dagger\,\psi_N \cdot \partial_\mu \bm{\pi} + \mathrm{h.c.} \,,
\label{eq:LagpiND}
\\
 \mathcal{L}_{\gamma N\Delta} &=&
  i\,e\,\bar{\psi}_\Delta^\mu\,T_3^\dagger \,\Gamma_{\mu\nu}\,\psi_N\,A^{\nu}
   + \mathrm{h.c.}\, ,
\label{eq:LaggaND}
\end{eqnarray}
where $\psi_\Delta^\mu$ is the Rarita-Schwinger $\Delta$ field operator, and
$\bm{T}$ is the $N \leftrightarrow \Delta$ isospin transition operator. The
$\pi N \Delta$ coupling constant $f_{\pi N \Delta}$ in Eq.~(\ref{eq:LagpiND})
is taken from the decay $\Delta \to \pi N$, which yields: $f_{\pi N \Delta}
\approx 1.95$. In Eq.~(\ref{eq:LaggaND}), the $\gamma N\Delta$ coupling
$\Gamma_{\mu\nu}$ has the form
\begin{equation} 
  \Gamma^{\mu\nu} = G_M\,\Gamma_M^{\mu\nu} + G_E\,\Gamma_E^{\mu\nu} \,,
\end{equation}
where $\Gamma_M^{\mu\nu}$ and $\Gamma_E^{\mu\nu}$ denote the magnetic and
electric $\gamma N \Delta$ vertices, respectively,
\begin{eqnarray}
\Gamma_M^{\mu\nu} &=&
  -\frac{3}{4 M_N} \, \frac{1}{(M_\Delta + M_N)} \,
  \varepsilon^{\mu\nu \kappa \lambda} \,
  ( p_\Delta + p)_\kappa \, k_\lambda \,,
\\
\Gamma_E^{\mu\nu} &=&
  -\Gamma_M^{\mu\nu} - \frac{3i\gamma_5}{(M_\Delta+M_N)\,(M_\Delta-M_N)^2\,M_N}
  \left(\varepsilon^{\mu \sigma \kappa \lambda} \,
  ( p_\Delta + p )_\kappa \, k_\lambda \right) \,
  \left(\varepsilon^{\nu}_{\ \sigma}{}^{\rho \tau}( p_\Delta )_\rho
  k_\tau \right)  \,.
\end{eqnarray}
The magnetic and electric $\gamma N \Delta$ couplings $G_M$ and
$G_E$ at the real photon point will be adjusted in the following.
\newline
\indent
Using the effective Lagrangians of Eqs.~(\ref{eq:LagpiND}) and
(\ref{eq:LaggaND}), we can write the $\Delta$ resonance
contribution of Fig.~\ref{fig:gap_piN_diag}(a) as follows~:
\begin{eqnarray} \label{eq:delpiN}
 v^\Delta_{\gamma\pi} &=& - e \, C_{\pi N} \, \frac{f_{\pi N \Delta}}{m_\pi} \,
     q^\alpha \varepsilon_\mu(k, \lambda)\, \bar u(p^{\prime}, s^{\prime}_N) \,
     \tilde{G}_{\alpha \beta}(p_\Delta) \, \big[ G_M \, \Gamma_M^{\beta \mu}
     \,+\, G_E \, \Gamma_E^{\beta \mu} \big] \, u(p, s_N) \,,
\end{eqnarray}
where $C_{\pi N}=2/3$ for $\gamma p \to \pi^0 p$ and $-\sqrt{2}/3$ for $\gamma
p \to \pi^+ n$, and $\tilde{G}_{\alpha\beta}$ is the $\Delta$ propagator given
by Eq.~(\ref{eq:redprop}).
\newline
\indent
To take account of the finite width of the $\Delta(1232)$ resonance, we
follow the procedure of
Refs.~\cite{DV01,ElAmiri:1992xa,LopezCastro:2000ep} by using a
complex pole description for the resonance excitation. This amounts to the
replacement
\begin{equation} \label{eq:cmplxmass}
  M_\Delta \,\longrightarrow\, M_\Delta - \frac{i}{2} \, \Gamma_\Delta \, ,
\end{equation}
in the propagator of Eq.~(\ref{eq:redprop}). This `complex mass scheme'
guarantees electromagnetic gauge invariance. In contrast, the use of a
Breit-Wigner propagator with an energy-dependent width will violate gauge
invariance when applied to the $\Delta$ contribution for the $\gamma p
\rightarrow \gamma \pi N$ reaction. For mass and width we take the complex pole
values given by the PDG~\cite{PDG02}: $(M_\Delta,~\Gamma_\Delta)$ = (1210,~100)
MeV, which provides a good description of the photoproduction data.
\newline
\indent
Since our main goal is to explore the role of the $\Delta^+(1232)$ MDM in the
$\gamma p \to \gamma \pi N$ reaction, it is not our purpose to precisely
reproduce the pion photoproduction data over a large energy range. For example,
the inclusion of an energy-dependence for the PS-PV mixing parameter and the
use of a Breit-Wigner distribution with an energy dependent $\Delta$ decay
width instead of the energy independent complex pole description can improve
the description of the pion photoproduction data. However, these improvements
would create problems in maintaining gauge invariance in the $\gamma p \to
\gamma \pi N$ reaction, as will be discussed in detail in Sec. IV. Rather, our
strategy is to determine the very small set of parameters within the model
outlined above, which gives a reasonable description of the pion
photoproduction in the $\Delta(1232)$ region, and then apply the same model to
the $\gamma p \to \gamma \pi N$ reaction.
%
\begin{figure}[h]
\includegraphics[width=0.35\columnwidth]{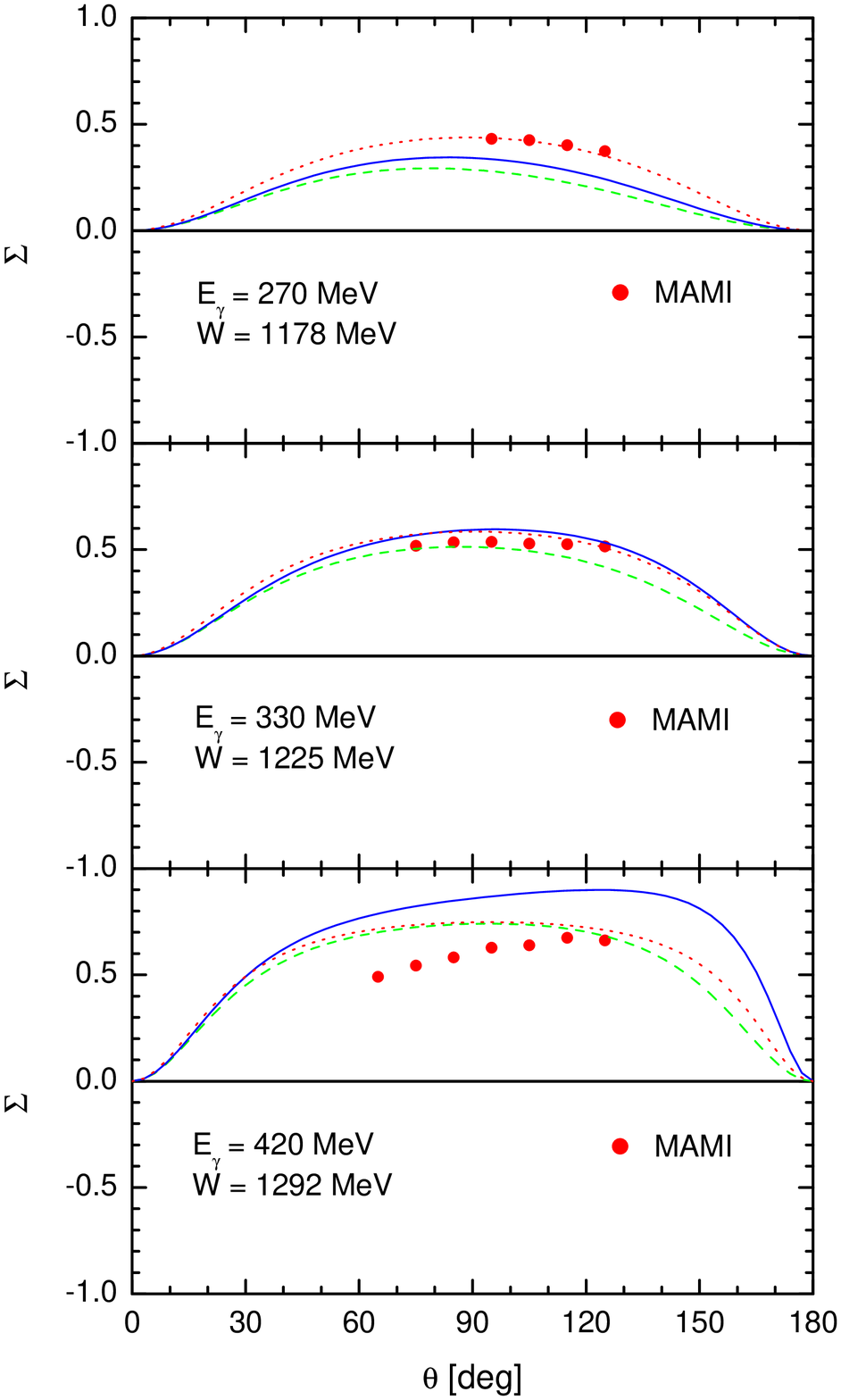}
\includegraphics[width=0.35\columnwidth]{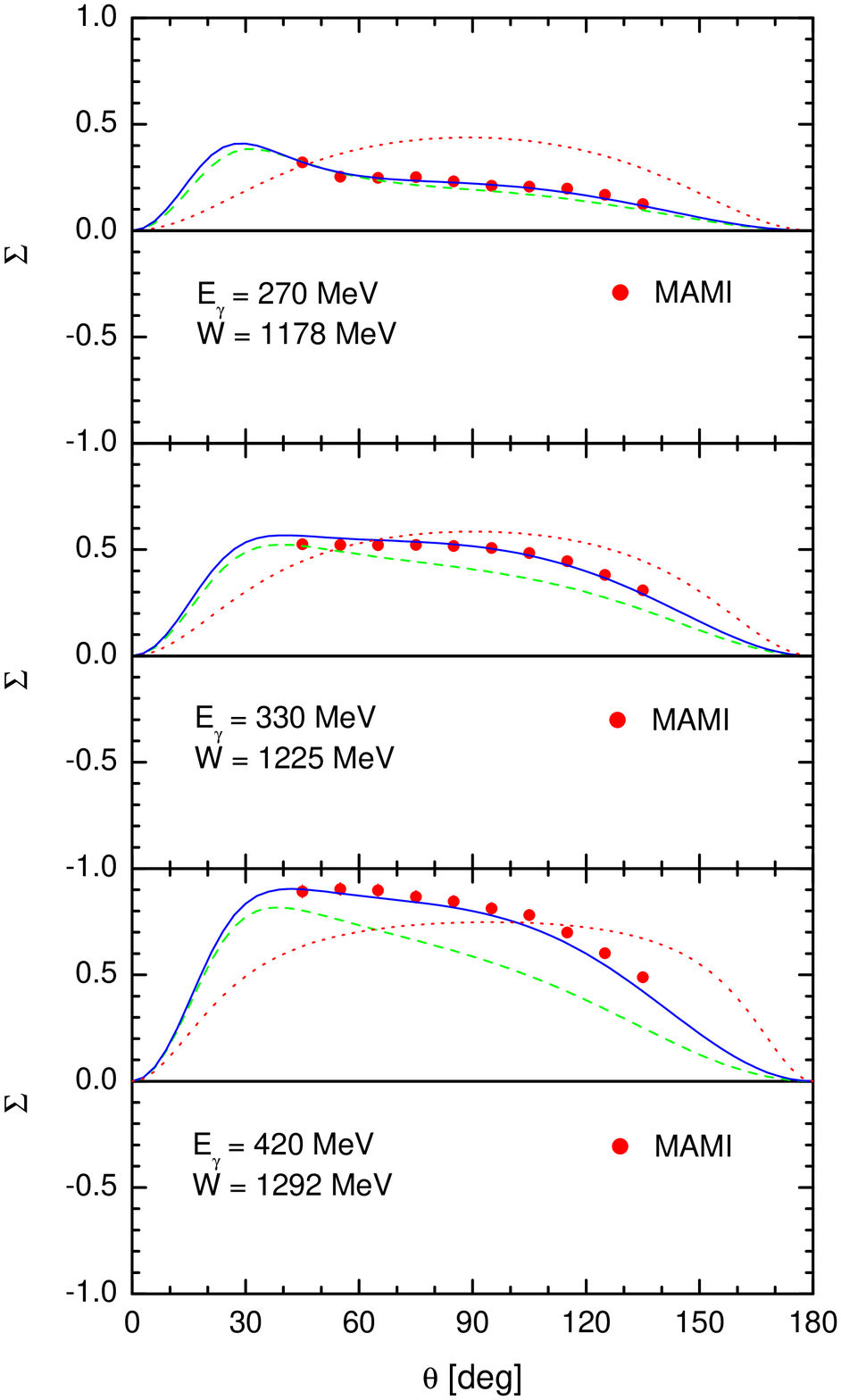}
\caption{\label{fig:gap_piop_beam}
Beam asymmetry for $\gamma p \rightarrow \pi^0 p$ (left panel)
and $\gamma p \rightarrow \pi^+ n$ (right panel)
at different photon {\it lab} energies $E_\gamma$
as function of the {\it c.m.} angle $\theta$.
The data are from MAMI~\cite{Beck:1999ge}. See
Fig.~\ref{fig:gap_piop_tot} for further notation. }
\end{figure}
%
\newline
\indent
In Fig.~\ref{fig:gap_piop_tot},
we show our results for the total $\gamma p \to \pi^0 p$ cross section with the
parameters $G_M = 3.00$ and $G_E = 0.065$, and compare with the data from
Refs.~\cite{McPherson:1964,Fissum:1996,MacCormick:1996jz,Ahrens:2000}. The
solid curve denotes the results obtained with our full unitary model while the
dotted curve indicates the unitarized $\Delta$ resonance contributions of
Eq.~(\ref{eq:Duni1}). If we approximate the t-matrix by the transition
potential of Eq.~(\ref{eq:vBD}) and replace $M_{\Delta}\to
M_{\Delta}-\frac{i}{2}\Gamma_{\Delta}$, we obtain the tree-level result
represented by the dashed curve.
This corresponds to the previous results of Ref.~\cite{DV01},
except that we use the above values of $G_M$ and $G_E$. We
find that the fully unitary model describes the total cross sections
for both the $\gamma p \to \pi^0 p$ and $\gamma p \to \pi^+ n$
reactions very well from threshold to $E^{\mathrm{lab}}_\gamma = $ 450~MeV,
as is shown in Fig.~\ref{fig:gap_piop_tot}.
The difference between the unitarized
result and the tree-level calculation indicates the size of the rescattering
effects, which turns out to be relatively large for the $\gamma p\to\pi^0 p$
reaction. We find that even though one can improve the description of the total
cross sections within the tree-level approximation by adjusting the model
parameters $G_M$ and $G_E$, it is not possible to achieve a satisfactory
tree-level description for the differential and polarization cross sections
discussed next.
\newline
\indent
The results for the differential cross sections for $\gamma p \to \pi^0 p$ are
shown in Fig.~\ref{fig:gap_piop_diff} (left panel).
They agree well with the data from
Refs.~\cite{Beck:1999ge,Genzel:1974} except for the highest energy
$E^{\mathrm{lab}}_\gamma = $ 420~MeV. The deviation may be traced back to the
fact that the $\Delta$ propagator of Eq.~(\ref{eq:redprop}) contains an
additional spin-1/2 component. In the case of charged pion photoproduction, the
angular distribution shows an interference pattern between background and
resonance contributions, which leads to an enhancement of backward production
below the resonance and a sharp rise in the forward direction above the
resonance.
\newline
\indent
The beam asymmetries for neutral and charged pion production are shown in
Fig.~\ref{fig:gap_piop_beam}. While we observe some deviations
between the model and the neutral pion data at the highest energy,
the process $\gamma p\to \pi^+ n$ is well described over the full
energy range. In particular we note a considerable improvement
of the angular distribution due to the unitarization effects.
%
\begin{figure}[h]
\includegraphics[width=0.42\columnwidth]{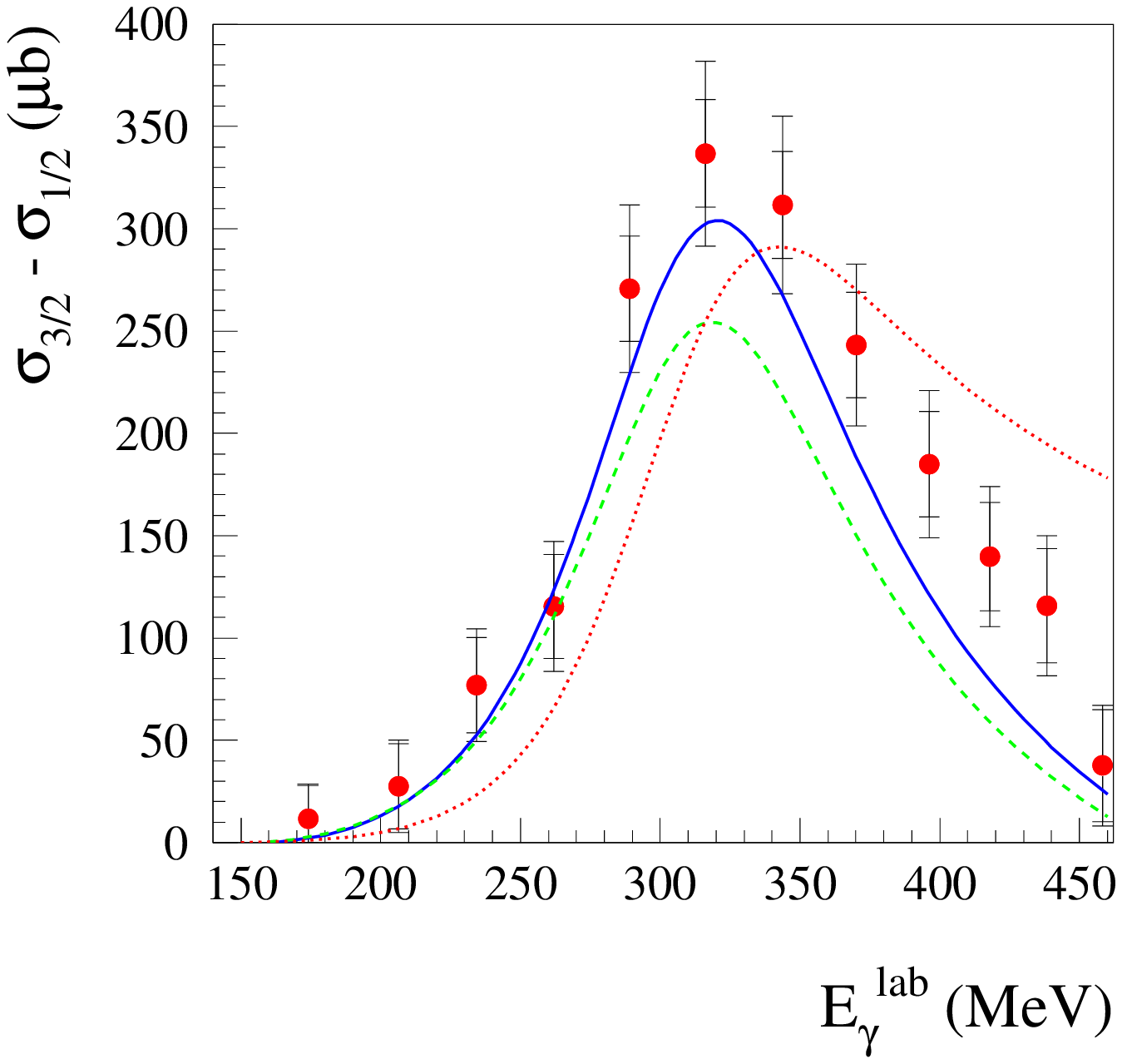}
\includegraphics[width=0.42\columnwidth]{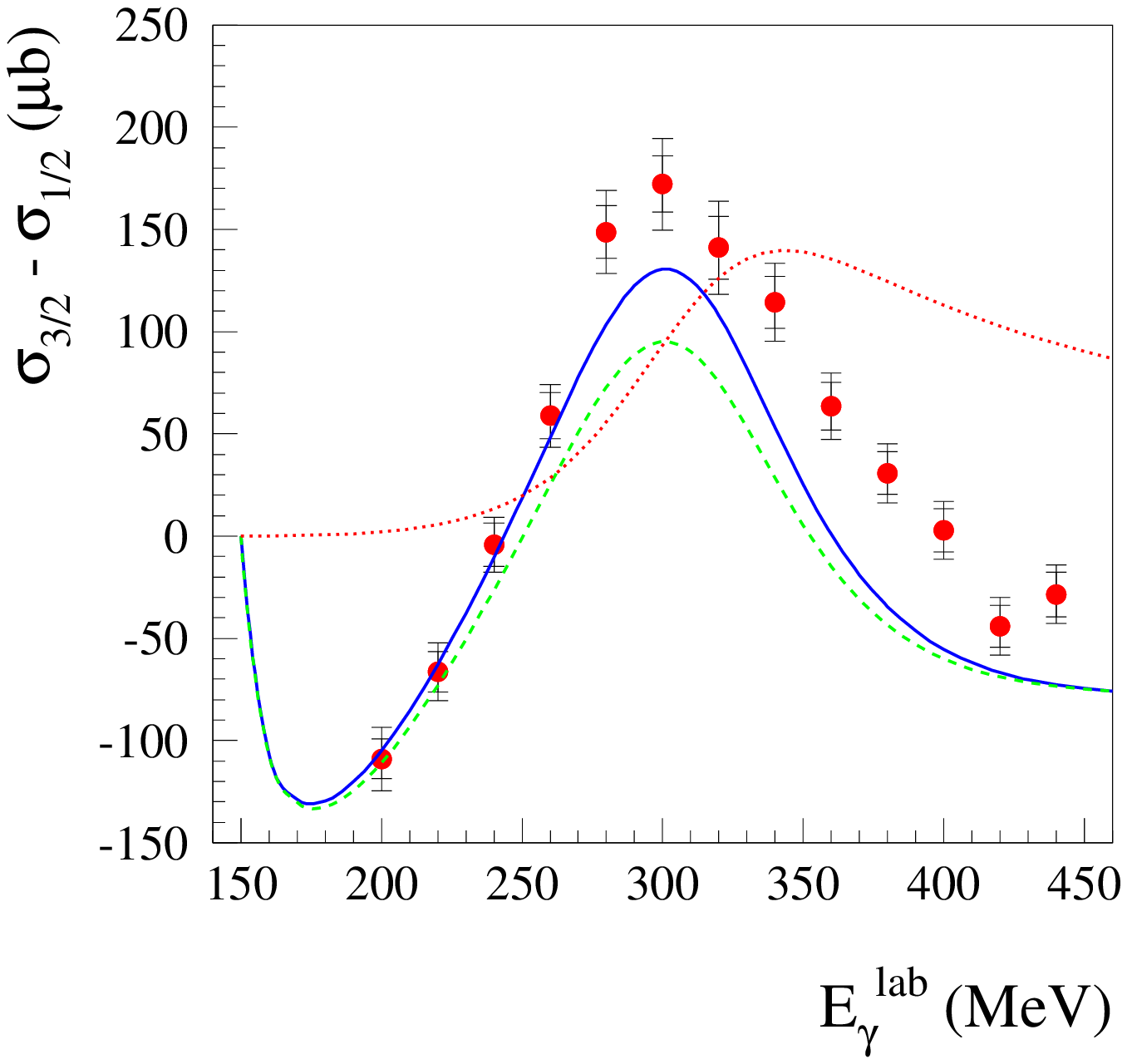}
\caption{\label{fig:gap_piop_hel}
Helicity cross section difference $\sigma_{3/2} - \sigma_{1/2}$
for the $\gamma p \rightarrow \pi^0 p$ reaction (left panel)
and the $\gamma p \rightarrow \pi^+ n$ reaction (right panel).
The data are from MAMI~\cite{Ahrens:2000} : inner error bars correspond with
statistical errors, outer error bars include systematical errors.
See Fig.~\ref{fig:gap_piop_tot} for further notation. }
\end{figure}
%
%
\begin{figure}[h]
\includegraphics[width=0.40\columnwidth]{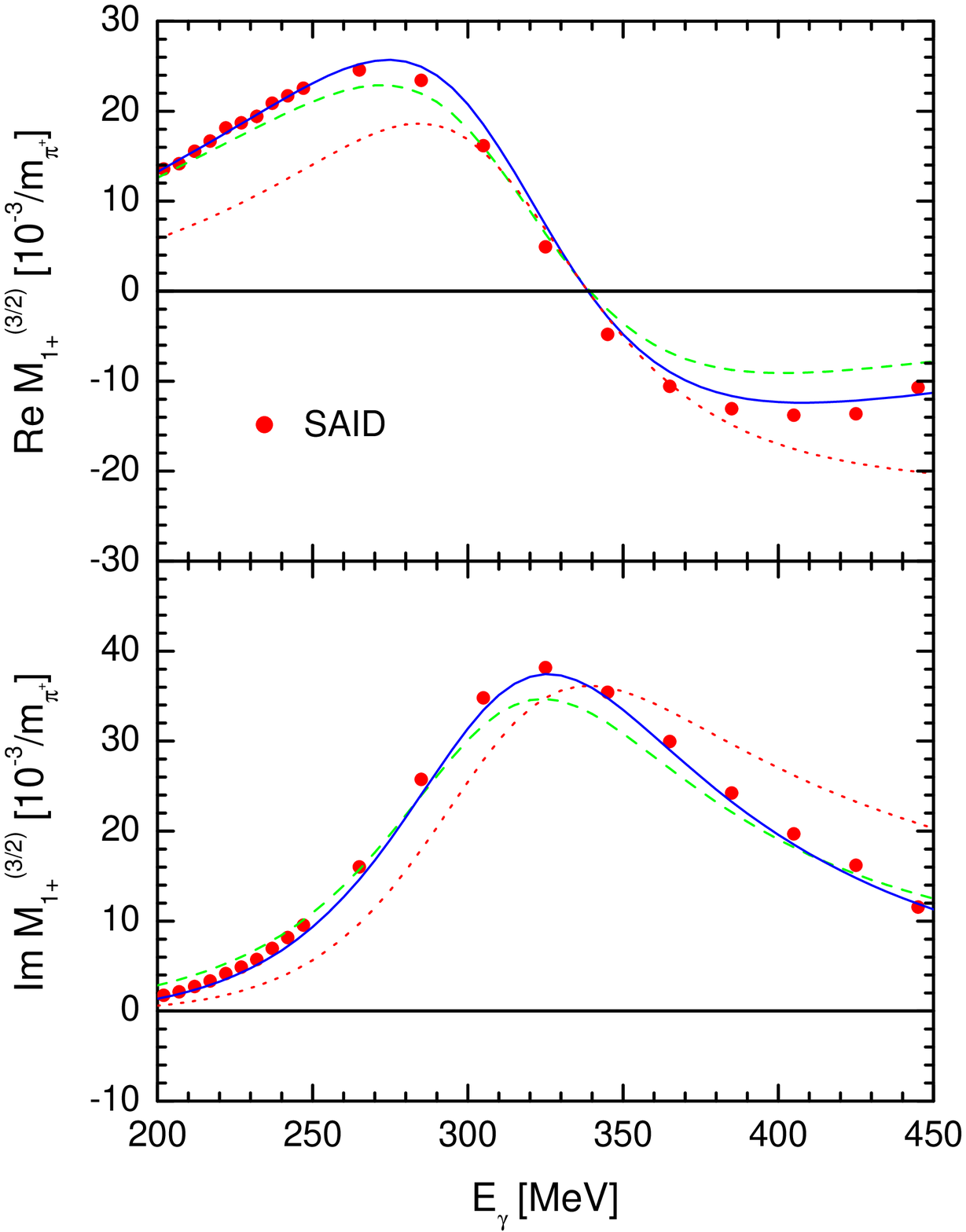}
\includegraphics[width=0.40\columnwidth]{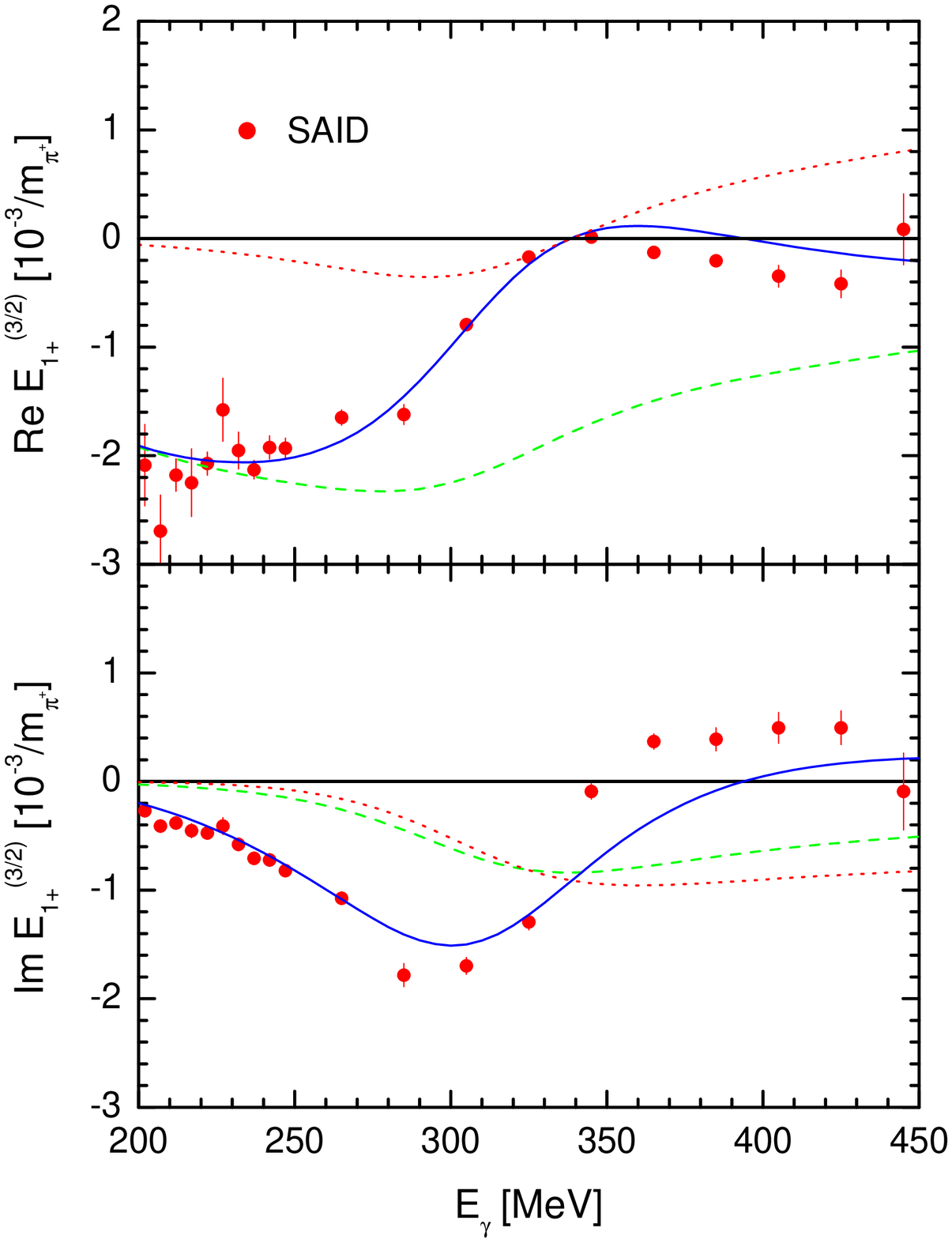}
\caption{\label{fig:gap_piN_M1p}Multipole amplitudes
$M_{1+}^{(3/2)}$ (left panel) and $E_{1+}^{(3/2)}$ (right panel)
for pion photoproduction as function of the photon {\it lab} energy
$E_\gamma$.
The data are taken from the SAID analysis~\cite{Arndt:2002xv}.
See Fig.~\ref{fig:gap_piop_tot} for further notation. }
\end{figure}
%
\newline
\indent
Recently, the helicity dependent total
cross sections $\sigma_{3/2}$ ($\sigma_{1/2}$)
for the absorption of circularly polarized photons on nucleons
in total helicity states of 3/2 (1/2) have been measured for the
$\gamma p \to \pi^0 p$ and $\gamma p \to \pi^+ n$
reactions \cite{Ahrens:2000}.
We show the comparison between our unitary model and these data in
Fig.~\ref{fig:gap_piop_hel}. For the $\gamma p \to \pi^0 p$ reaction
the $\sigma_{3/2}$ cross section is dominated by the
$\Delta(1232)$ resonance,
which yields positive values of $\sigma_{3/2} - \sigma_{1/2}$
over the full energy region. For the $\gamma p \to \pi^+ n$ channel,
however, the strong non-resonant pion production leads to
a large $\sigma_{1/2}$ cross section near threshold followed by
$\sigma_{3/2}$ dominance in the $\Delta(1232)$ region.
Altogether we obtain a qualitatively
good description of the helicity dependent cross sections
for both $\gamma p \to \pi^0 p$ and $\gamma p \to \pi^+ n$,
with some deviations appearing on the high energy side of the $\Delta(1232)$
region.
\newline
\indent
Fig.~\ref{fig:gap_piN_M1p} shows the multipole
amplitudes in the $\Delta$ channel, $M_{1+}$ and $E_{1+}$, obtained with our
best fit values $G_M=3.0$ and $G_E=0.065$.
These results are compared with the
SM02 solution of the SAID partial wave analysis~\cite{Arndt:2002xv}. The
unitarized model reproduces the dominant $M_{1+}^{(3/2)}$ multipoles from SAID
quite well, whereas the agreement is less perfect for the $E_{1+}^{(3/2)}$
multipole. Since the latter multipole is small ($G_E / G_M \sim 2\%$), these
deviations are however of little importance for our conclusions. As has been
stated earlier, the inclusion of higher resonances such as the $P_{11}(1440)$
could improve our calculation at the larger energies. However, a consistent
description of these higher resonances is outside the scope of our current
study and therefore left for future work.


\section{Unitary model for the $\gamma p \to \gamma \pi N$ reaction}
\label{sec:gapi}

In this section we extend the previously constructed model for pion
photoproduction to the reaction $\gamma p \to \gamma \pi N$ in the
$\Delta(1232)$ resonance region, which will then be used as a tool to
investigate the size of the $\Delta(1232)$ MDM.
After discussing the tree level processes for the
$\gamma p \to \gamma \pi N$ reaction, we will subsequently extend this
description and present a unitary model for the $\gamma p \to \gamma \pi^0 p$
reaction in the $\Delta(1232)$ region.

\begin{figure}
\centering
\includegraphics[width=0.8\columnwidth]{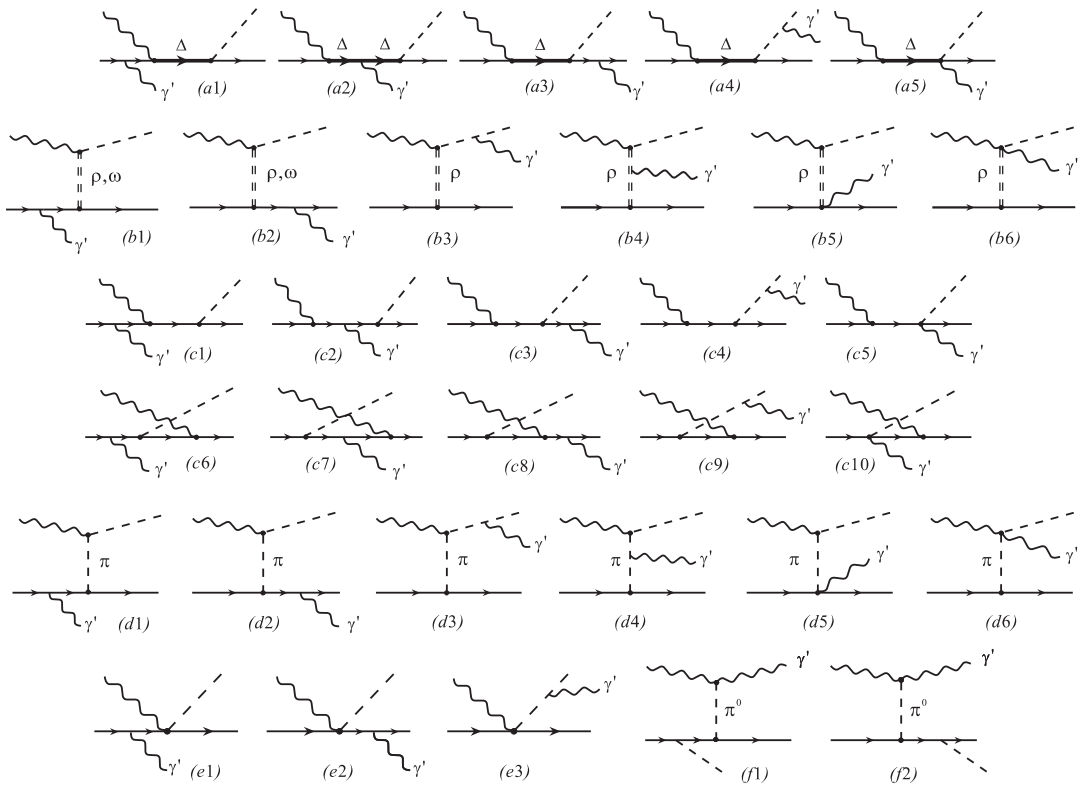}
\caption{\label{fig:gap_gapiN_diag}Tree diagrams considered in the calculation
of the $\gamma p \rightarrow \gamma \pi N$ reaction in the $\Delta(1232)$
region
: $\Delta$ resonance (a1-a5), vector-meson exchange (b1-b6), nucleon-pole
(c1-c10), pion-pole (d1-d6), Kroll-Rudermann (e1-e3), and anomaly diagrams
(f1-f2). }
\end{figure}

We start from the tree diagrams in Fig.~\ref{fig:gap_piN_diag} as prescribed by
the effective Lagrangian for $\gamma p \to  \pi  N$ and couple a photon to all
the charged particles. The resulting diagrams are shown in
Fig.~\ref{fig:gap_gapiN_diag}. The diagrams Fig.~\ref{fig:gap_gapiN_diag}(b-e)
referring to vector-meson exchanges, nucleon and pion pole as well as
Kroll-Ruderman terms can be evaluated with the interaction Lagrangians given in
Sec.~III or by minimal substitution of pion-nucleon Lagrangians given in that
section. For example, the $\gamma\pi NN$  and $\gamma\gamma\pi\pi$ vertices can
be obtained by replacing the derivative $\partial^\mu$ in $\mathcal{L}_{\pi
NN}$ and $\mathcal{L}_{\gamma\pi\pi}$ by the covariant derivative $\partial^\mu
+ i Q A^\mu$, where $Q$ is the charge of the respective pion. By construction
this set of diagrams is therefore gauge invariant by itself
with respect to both initial and final photons.
In Fig.~\ref{fig:gap_piN_diag}, we display the diagrams for both the
$\gamma p \to \gamma \pi^0 p$ and the $\gamma p \to \gamma \pi^+ n$
reaction. When omitting the coupling
to the pion lines and the contact diagrams, diagrams (a - c) yield
the tree diagrams considered in Ref.~\cite{DV01}.
In addition to Ref.~\cite{DV01}, we include the diagrams
Fig.~\ref{fig:gap_gapiN_diag}(f) resulting from the $\pi^0 \to \gamma \gamma$
anomaly as given by the Wess-Zumino-Witten Lagrangian~\cite{Wess71},
\begin{eqnarray} \label{eq:anomaly}
L_{WZW} = \frac {\alpha_{em}}{8\pi F_\pi} \epsilon_{\mu\nu\alpha\beta}
F^{\mu\nu}F^{\alpha\beta} \pi^0\,,
\end{eqnarray}
with $F^{\mu\nu} = \partial^\mu A^\nu -\partial^\nu A^\mu,$ and
$F_\pi=92.4$~MeV the pion decay constant.
Note that in the soft photon limit for the final photon, i.e. $k' \to 0$,
the anomaly diagrams vanish as they are linear in the final photon momentum
$k'$.
\newline
\indent
The $\Delta$ resonance diagrams of
Fig.~\ref{fig:gap_gapiN_diag}(a) can be similarly evaluated by use of the
previously described Lagrangians except for
diagram~\ref{fig:gap_gapiN_diag}(a2). The latter diagram contains the
interaction Lagrangian
\begin{eqnarray} \label{eq:GDD}
L_{\gamma\Delta\Delta} =  e_\Delta \bar{\psi}_{\Delta}^{\beta'} \left\{
g_{\beta'\beta}\,\left(\gamma_\nu A^\nu
-\frac{\kappa_\Delta}{2 M_\Delta} \sigma_{\nu\lambda}\,
\partial^\lambda A^\nu \right)\,+\,\frac{1}{3}\,
\left(\gamma_\beta\gamma_\nu\gamma_{\beta'}\,-\,\gamma_{\beta}
g_{\nu\beta'}\,-\,\gamma_{\beta'}
g_{\nu\beta}\right) A^\nu \right\}\,\psi_\Delta^\beta \, ,
\end{eqnarray}
which contains the information on $\kappa_\Delta$, the MDM of the
$\Delta(1232)$ resonance. Therefore, in comparison with the $\gamma p \to \pi
N$ process, the only new parameter entering in the description of the $\gamma p
\rightarrow \gamma \pi N$ process is the $\Delta^+(1232)$ anomalous magnetic
moment $\kappa_{\Delta^+(1232)}$.
The $\gamma \Delta \Delta$ vertex of Eq.~(\ref{eq:GDD}) satisfies the
electromagnetic Ward identity with the $\Delta$ propagator of
Eq.~(\ref{eq:redprop}). Gauge invariance is preserved
when using the complex pole description of Eq.~(\ref{eq:cmplxmass}).
\newline
\indent
In principle, the $\gamma \Delta \Delta$ vertex and hence
$\kappa_\Delta$ is a function of $k'^2$, $p_{\Delta}^2$,
$p_{\Delta}^{\prime \, 2}$, the four-momentum squared of the emitted photon,
initial and final $\Delta$, respectively.
As we are studying the transition induced by real photons,
$k'^2 = 0$. If we restrict ourself to the
$\Delta(1232)$ resonance region,
then we can choose $p_{\Delta}^2 = M_\Delta^2$,
and $\kappa_{\Delta}$ will depend only on $p_{\Delta}^{\prime 2}$.
In the soft-photon limit, we will further have
$p_\Delta^{\prime 2} = M_\Delta^2$.
By assuming that $\kappa_\Delta$ is a slowly varying function of
$p_\Delta^{\prime 2}$, we can then treat  $\kappa_\Delta$
as a constant in the soft-photon region. We will confine ourselve
to this kinematical region in our current investigation.
%
\begin{figure}[tbp]
\begin{center}
\includegraphics[width=.95\columnwidth]{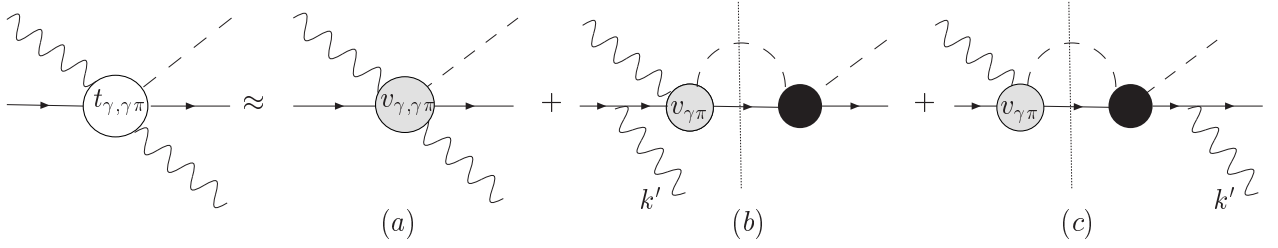}
\end{center}
\caption{Model for the $T$-matrix for the $\gamma p \to \gamma \pi^0 p$
reaction used in this work.
The transition potential $v_{\gamma, \gamma \pi}$ (diagram a) corresponds
with the diagrams of Fig.~\ref{fig:gap_gapiN_diag}.
The rescattering contributions (diagrams b and c) are
evaluated in the soft-photon approximation for the final photon,
i.e. $k' \to 0$.
The transition potential $v_{\gamma \pi}$ corresponds with the
diagrams of Fig.~\ref{fig:gap_piN_diag}.
The black blob corresponds with the full $T$-matrix $t_{\pi N}$ for $\pi N$
scattering. The vertical dotted lines indicate that the $\pi N$ intermediate
state is taken on-shell ($K$-matrix approximation).}
\label{fig:rescatter}
\end{figure}
%
\newline
\indent
We next turn to the rescattering contribution for the
$\gamma p \to \gamma \pi^0 p$ reaction. In this work, we estimate
this rescattering in the $K$-matrix approximation, i.e. considering
only on-shell rescattering. Furthermore, in the soft-photon
limit for the final photon, the $T$-matrix for the
$\gamma p \to \gamma \pi^0 p$ reaction has to be directly proportional
to the full $T$-matrix for the $\gamma p \to \pi^0 p$ reaction
as previously constructed (see Fig.~\ref{fig1}).
We construct the full $T$-matrix for
$\gamma p \to \gamma \pi^0 p$ as shown in Fig.~\ref{fig:rescatter}
by the expression
\begin{eqnarray}
\label{eq:tmatrix_gagapi}
t_{\gamma,\gamma\pi} (\bm{k'} \bm{q}, W_{\pi N} ; \bm{k}, \sqrt{s} )
\approx v_{\gamma,\gamma\pi}(\bm{k'} \bm{q}; \bm{k} \bm) +
e \, \varepsilon^*_\nu(k', \lambda') \,
\left( \frac{p'^\nu}{p' \cdot k'} - \frac{p^\nu}{p \cdot k'} \right) \,
\left[ t_{\gamma \pi}(\bm{q}, \bm{k}; \sqrt{s})
- v_{\gamma \pi}(\bm{q}, \bm{k}) \right].
\end{eqnarray}
The first term in Eq.~(\ref{eq:tmatrix_gagapi}), denoted by the
transition potential $v_{\gamma,\gamma\pi}$,
is the sum of all tree diagrams shown in Fig.~\ref{fig:gap_gapiN_diag}.
As discussed before, this term is gauge invariant by itself with respect
to both intial and final photons.
In the soft-photon limit (i.e. $k' \to 0$), it reduces to~:
\begin{eqnarray}
v_{\gamma,\gamma\pi}(\bm{k'} \bm{q}; \bm{k} \bm)
\stackrel{k' \to 0}{\longrightarrow}
e \, \varepsilon^*_\nu(k', \lambda') \,
\left( \frac{p'^\nu}{p' \cdot k'} - \frac{p^\nu}{p \cdot k'} \right) \,
v_{\gamma \pi}(\bm{q}, \bm{k}),\
\label{eq:soft_gagapi1}
\end{eqnarray}
where $v_{\gamma \pi}$ is the transition potential for the
$\gamma p \to \pi^0 p$ reaction as shown in Fig.~\ref{fig:gap_piN_diag}.
The second term in Eq.~(\ref{eq:tmatrix_gagapi})
is the rescattering contribution. Since we only keep the leading
term in the outgoing photon energy for the rescattering term ($k' \to 0$),
this amounts to evaluate the $T$-matrix $t_{\gamma \pi}$
in the second term of Eq.~(\ref{eq:tmatrix_gagapi})
in soft-photon kinematics, i.e. with
nucleon momenta $\bm{p} = - \bm{k}$, $\bm{p'} = - \bm{q}$,
and  at total energy $\sqrt{s}$, with $s = (k + p)^2$,
as we work in the {\it c.m.} system of the $\gamma p \to \gamma \pi N$
reaction.
For the $t$-matrix $t_{\gamma \pi}$, we adopt the unitary model as
discussed in section~\ref{sec:pi}.
Evaluating the rescattering term for the $\gamma p \to \gamma \pi N$
process in the soft-photon limit, as done in Eq.~(\ref{eq:tmatrix_gagapi}),
ensures us that the full amplitude $t_{\gamma, \gamma \pi}$
satisfies the low energy theorem.
Indeed, using Eq.~(\ref{eq:soft_gagapi1}),
we immediately verify from Eq.~(\ref{eq:tmatrix_gagapi}) that
\begin{eqnarray}
t_{\gamma,\gamma\pi} (\bm{k'} \bm{q}, W_{\pi N} ; \bm{k}, \sqrt{s} )
\stackrel{k' \to 0}{\longrightarrow}
e \, \varepsilon^*_\nu(k', \lambda') \,
\left( \frac{p'^\nu}{p' \cdot k'} - \frac{p^\nu}{p \cdot k'} \right) \,
t_{\gamma \pi}(\bm{q}, \bm{k}; \sqrt{s}),
\label{eq:soft_gagapi2}
\end{eqnarray}
as required by the low energy theorem.
\newline
\indent
Furthermore, both terms in
our model for the full $T$-matrix $t_{\gamma, \gamma \pi}$ in
Eq.~(\ref{eq:tmatrix_gagapi}) satisfy gauge invariance with
respect to both intial and final photons.
One notices in particular that the rescattering contribution
( second term in Eq.~(\ref{eq:tmatrix_gagapi}) )
is by construction gauge invariant
with respect to the final photon
(as is evident when replacing $\varepsilon^*(k', \lambda')$ by $k'$).
We further point out that the rescattering contributions
of Fig.~\ref{fig:rescatter} (b) and (c) are obtained by summing over both
$\pi^+ n$ and $\pi^0 p$ intermediate states in the loop.
\newline
\indent
To evaluate the rescattering contribution beyond the soft-photon limit
is much more complicated and requires a
coupled channel calculation for both
$\gamma N \to \gamma \pi N$ and $\pi N \to \gamma \pi N$ processes,
because the outgoing photon can be produced not only in the initial step
but also by a pion while rescattering off the nucleon.
Furthermore, the exact conservation of gauge-invariance in such an approach
requires to introduce vertex corrections wherever the photon is emitted
between two different pion rescatterings.
We leave such a description to a future work, because our
evaluation of the rescattering effects
for the $\gamma p \to \gamma \pi^0 p$ process is motivated by the
experimental situation where one stays relatively close to the
soft-photon limit, i.e., at outgoing photon energies up to about 100 MeV.
Furthermore, for the tree level contribution $v_{\gamma, \gamma \pi}$ in
Eq.~(\ref{eq:tmatrix_gagapi}) we do not make
a soft-photon approximation, but calculate the full outgoing photon energy
dependence.
Because the effect of the rescattering turns out to be modest in
all the following calculations it is a reasonable approximation
to evaluate the rescattering contribution
in the soft-photon limit for those kinematics.

\section{Results and Discussion for $\gamma p\to\gamma\pi N$ Observables}
\label{sec:results}

%
\begin{figure}[h]
\includegraphics[width=0.45\columnwidth]{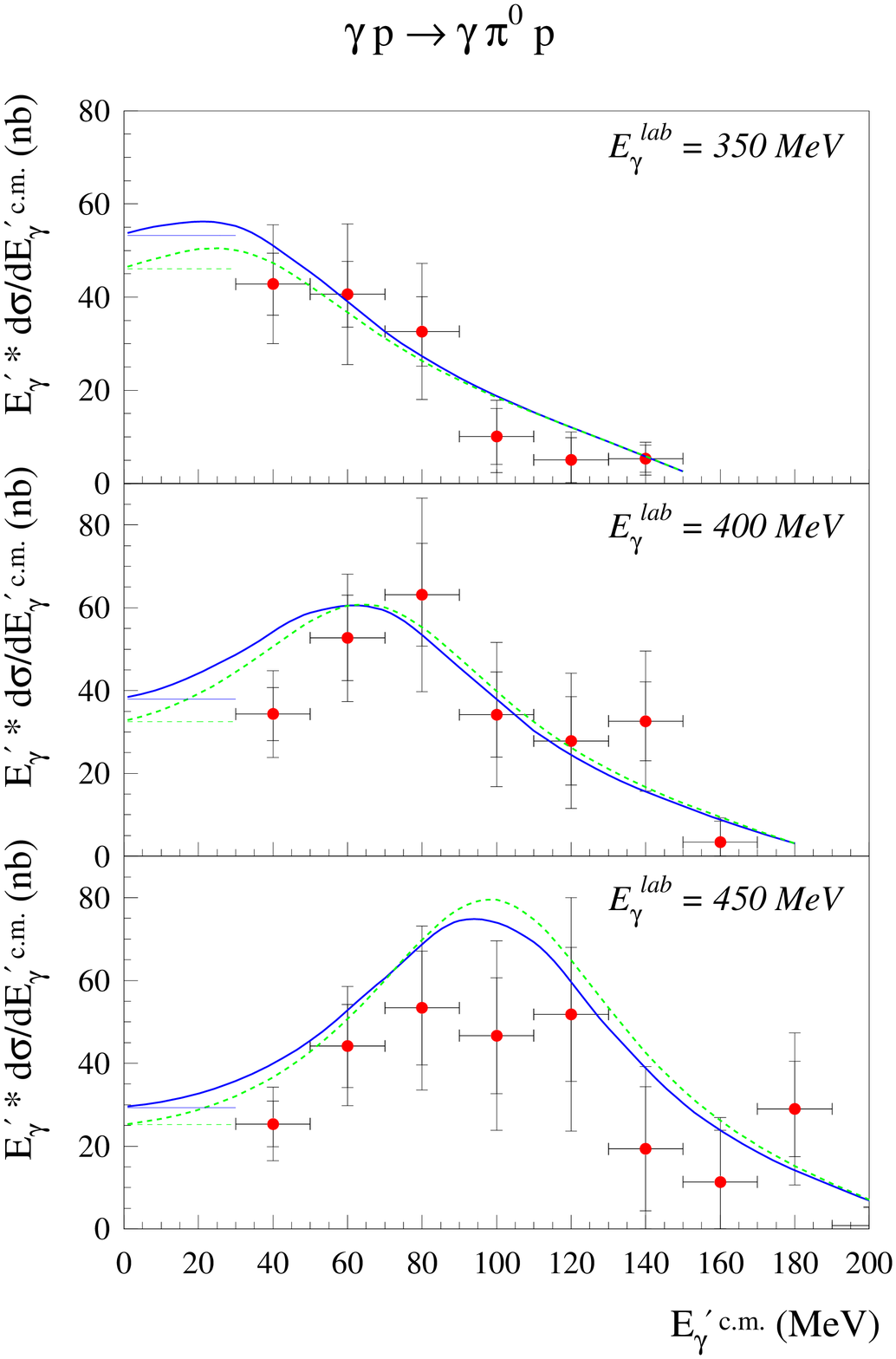}
\includegraphics[width=0.45\columnwidth]{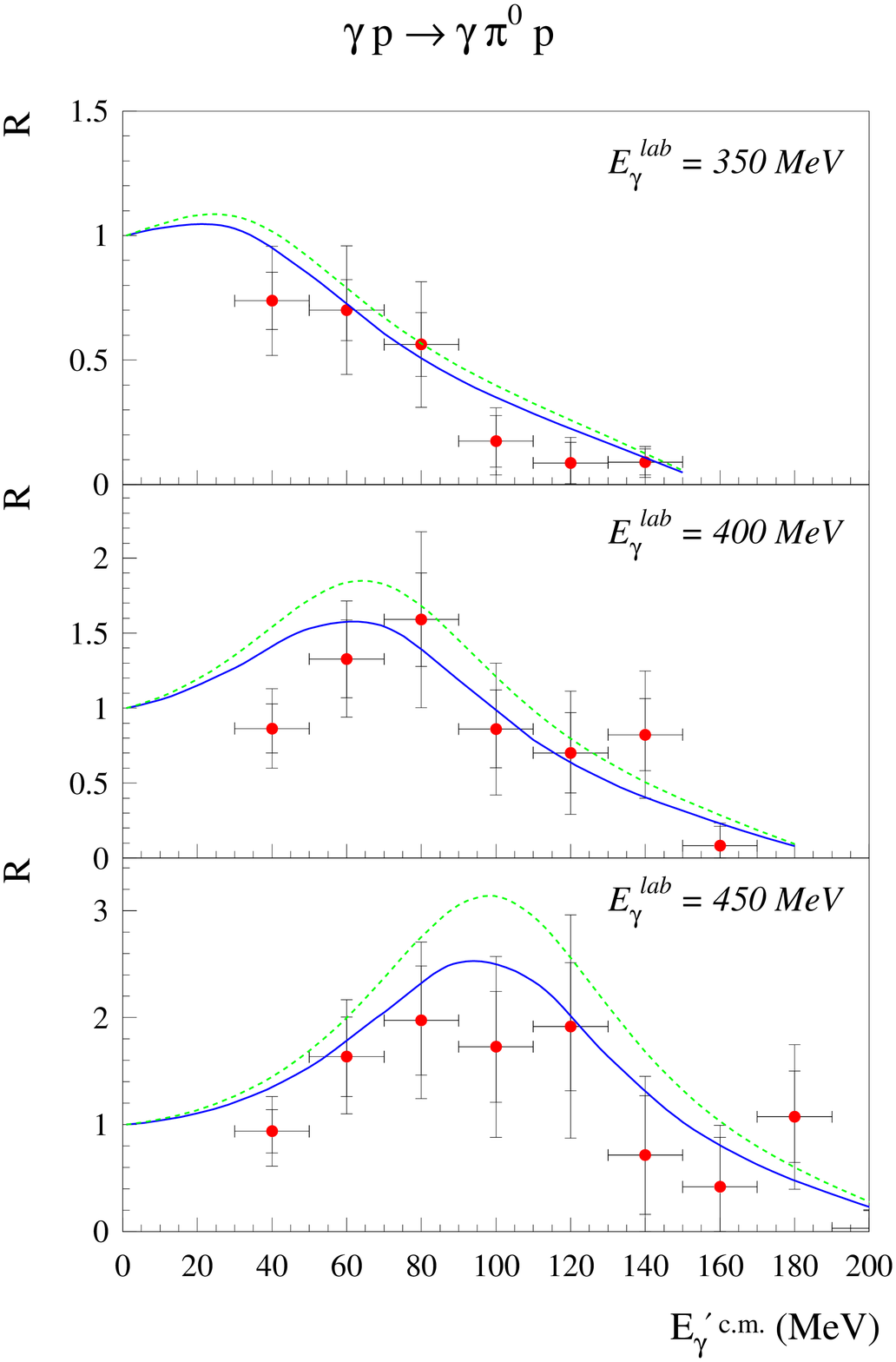}
\caption{\label{fig:gapiop_tcs}
Left panel : outgoing photon energy dependence of the cross
section $d\sigma/dE'^{c.m.}_\gamma$ multiplied by $E'^{c.m.}_\gamma$
for the $\gamma p \rightarrow \gamma \pi^0 p$ reaction.
A comparison is shown between the tree level calculation (dashed curves),
and the result of the unitary model (solid curves).
All results are obtained with $\kappa_{\Delta^+} = 3$.
The horizontal curves at small values of $E'^{c.m.}_\gamma$ are
obtained using the low energy theorem
for the tree level model (thin dashed curves)
and the unitary model (thin solid curves).
Right panel : ratio $R$, as defined in Eq.~(\ref{eq:R1}), for the $\gamma
p \rightarrow \gamma \pi^0 p$ reaction and with the same conventions as
on left panel.
The data in both panels
are from MAMI~\cite{Kot02} : inner error bars correspond to
statistical errors, outer error bars include systematical errors.
}
\end{figure}
%
In Fig.~\ref{fig:gapiop_tcs}, we show the outgoing
photon energy dependence of the cross section
$d\sigma/dE'_\gamma$ for the $\gamma p \rightarrow \gamma \pi^0 p$
reaction integrated over the
photon and pion angles for three incoming photon energies through the
$\Delta(1232)$ region.
Because the cross sections exhibit the characteristic
bremsstrahlung behavior, i.e. dropping as $1/E'_\gamma$
at low energies $E'_\gamma$, we display the cross sections in
the left panel of Fig.~\ref{fig:gapiop_tcs} multiplied by $E'_\gamma$.
\newline
\indent
In the soft photon limit ($E'_\gamma \rightarrow 0$),
gauge invariance provides a
model-independent relation between the cross sections for the $\gamma p
\rightarrow \gamma \pi N$ and $\gamma p \rightarrow \pi N$ reactions. This
low-energy theorem was derived in Ref.~\cite{Chiang:2002ah} for radiative
photoproduction of a neutral meson. In Appendix~\ref{app:soft},
we first apply this theorem to the $\gamma p \to \gamma \pi^0 p$ reaction and
then extend it to the $\gamma p \rightarrow \gamma \pi^+ n$ process.
Its derivation is based on the observation
that in the soft-photon limit the $\gamma p \rightarrow \gamma \pi N$ reaction
is completely determined
by the bremsstrahlung process from the initial and final protons.
In this limit, when integrating the five-fold differential cross section of
Eq.~(\ref{eq:crosssection}) over the outgoing photon angles, we obtain the
three-fold differential cross section for the
$\gamma p \rightarrow \gamma \pi N$ process, which reduces in the
soft-photon limit to
\begin{equation} \label{eq:softint1}
 \left( \frac{d\sigma}{dE'_\gamma \, d \Omega_\pi} \right)^{c.m.}
\stackrel{E'_\gamma \to 0}{\longrightarrow}
  \frac{1}{E'_\gamma} \cdot
  \frac{e^2}{2 \pi^2} \, \cdot \, W(v) \,
  \left(\frac{d\sigma}{d\Omega_\pi}\right)^{c.m.} \,,
\end{equation}
where $(d\sigma/d\Omega_\pi)^{c.m.}$ is the differential cross
section for the $\gamma p \rightarrow \pi N$ process. The form of the
angular weight-function $W(v)$ is derived in Appendix~\ref{sec:LET1} as~:
\begin{equation}
\label{eq:wsoft}
W(v) = -1 \,+\, \left( \frac{v^2 + 1}{2 v} \right) \, \cdot \,
  \ln \left( \frac{v + 1}{v - 1} \right)  \,,
\end{equation}
with $v \equiv \sqrt{ 1 + 4 M_N ^2 / (-t) }$
and $t = (p' - p)^2$.
When integrating the five-fold
differential cross section of Eq.~(\ref{eq:crosssection})
over both the outgoing photon and the meson angles, we
obtain the following energy distribution for the
$\gamma p \rightarrow \gamma \pi N$ process~:
\begin{equation}
\label{eq:softint2}
  \left( \frac{d\sigma}{dE'_\gamma} \right)^{c.m.}
  \equiv
  \int d\Omega^{c.m.}_\pi
 \left( \frac{d\sigma}{dE'_\gamma d\Omega_\pi} \right)^{c.m.}
 \stackrel{E'_\gamma \to 0}{\longrightarrow}
  \frac{1}{E'_\gamma} \cdot \sigma_\pi \,,
\end{equation}
with a ``weight-averaged'' total cross section $\sigma_\pi$ for the $\gamma p
\rightarrow \pi N$ reaction,
\begin{equation} \label{eq:getaspl}
  \sigma_\pi
  \equiv
  \frac{e^2}{2 \pi^2} \int d\Omega^{c.m.}_\pi\, W(v)
  \left(\frac{d\sigma}{d\Omega_\pi}\right)^{c.m.}
  \,.
\end{equation}
The low energy theorem of Eq.~(\ref{eq:softint2}) provides a check for both
theoretical model calculations and experimental measurements, because
\begin{eqnarray}
\label{eq:R1}
  R \,\equiv \, \frac{1}{\sigma_\pi} \cdot
  E'_\gamma
  \frac{d\sigma}{dE'_\gamma}\rightarrow 1\quad
  {\mbox{for}}\ E'_\gamma \to 0\, .
\end{eqnarray}
\newline
\indent
At small values of $E'_\gamma$, one readily observes from
Fig.~\ref{fig:gapiop_tcs} (left panel)
that our theoretical calculation for the product
$E'_\gamma \cdot d\sigma/dE'_\gamma$ approaches a constant.
Because we model the $\gamma p \to \pi^0 p$
and $\gamma p \to \gamma \pi^0 p$ reactions within the same framework,
the low energy theorem is exactly satisfied, as follows from
Eq.~(\ref{eq:soft_gagapi1}) for the tree-level model and
Eq.~(\ref{eq:soft_gagapi2}) for the unitary model.
In the right panel of Fig.~\ref{fig:gapiop_tcs}, we show
the ratio $R$ constructed from our theoretical calculations of the $\gamma
p \rightarrow \gamma \pi^0 p$ and $\gamma p \rightarrow \pi^0 p$ reactions,
and compare with the data of Ref.~\cite{Kot02} for this same ratio,
where $\sigma_\pi$ is evaluated from the $\gamma p \to \pi^0 p$ data
using Eq.~(\ref{eq:R1}).
The first data for the
$\gamma p \to \gamma \pi^0 p$ process of Ref.~\cite{Kot02} show a clear
deviation from the soft-photon limit value $R = 1$
with increasing values of $E'_\gamma$.
One sees from Fig.~\ref{fig:gapiop_tcs} that our unitary model
gives a good overall description of the $E'_\gamma$ dependence of the
$\gamma p \to \gamma \pi^0 p$ reaction throughout the $\Delta$-region.
Compared with the tree-level model developed in Ref.~\cite{DV01},
our unitary model reduces the cross section at larger values of $E'_\gamma$,
and thus provides an improved description of the data.
%
\begin{figure}[h]
\includegraphics[width=0.5\columnwidth]{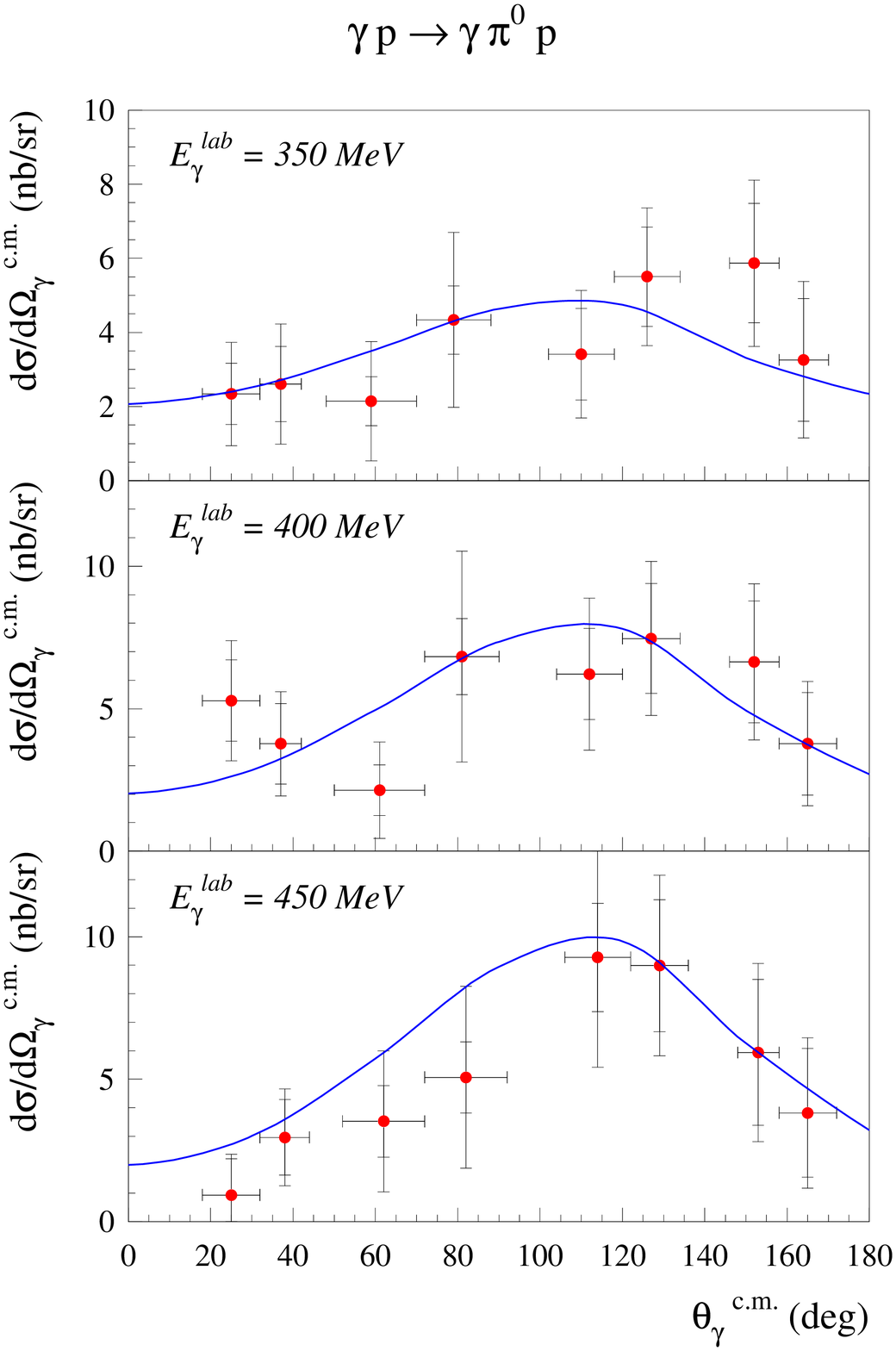}
\caption{
\label{fig:gap_gapiop_ang}
The angular distribution of the emitted photons for the
$\gamma p\to \gamma \pi^0 p$ {\it c.m.} cross section
$d\sigma/d\Omega_\gamma^{c.m.}$. The cross section is integrated over
the pion angles and over the outgoing photon energy range
$E'^{c.m.}_\gamma > 30$~MeV.
The results of the unitary model are shown with $\kappa_{\Delta^+} = 3$.
The data are from MAMI~\cite{Kot02} : inner error bars correspond to
statistical errors, outer error bars include systematical errors.
}
\end{figure}
%
\newline
\indent In Fig.~\ref{fig:gap_gapiop_ang}, we show the outgoing photon angular
dependence of the {\it c.m.} cross section $d \sigma / d \Omega'_\gamma$ for
the $\gamma p \to \gamma \pi^0 p$ reaction, which has also been measured in
Ref.~\cite{Kot02}. To compare with these data the cross section is integrated
over the pion angles and over the outgoing photon energy range
$E'^{c.m.}_\gamma > 30$~MeV. Our model reproduces the angular dependence of the
existing data, within their accuracy, rather well. We see that the model gives
a rather flat angular distribution at $E_\gamma^{lab} = 350$ MeV. At higher
incident photon energies, it displays a broad peak around photon {\it c.m.}
angles of 110$^o$. Such a structure is due to the interference between the
bremsstrahlung and $\Delta$-resonant mechanisms. Note that a pure
$\Delta$-resonant mechanism would yield a photon angular distribution peaked
around a {\it c.m.} angle of 90$^o$. Comparing the unitary model with the data
presented in Figs.~\ref{fig:gapiop_tcs} and \ref{fig:gap_gapiop_ang}, we
conclude that both the outgoing photon energy and angular distributions of the
$\gamma p \to \gamma \pi^0 p$ reaction through the $\Delta$ region show clear
deviations from a pure bremsstrahlung dominated process as obtained in the
soft-photon limit.
%
\begin{figure}[h]
\includegraphics[width=0.5\columnwidth]{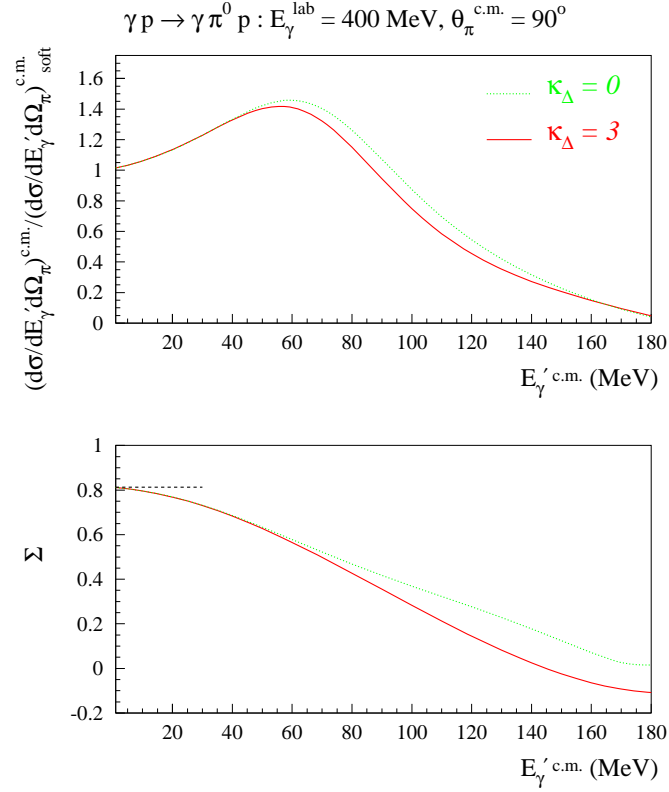}
\caption{
\label{fig:gap_gapiop_3fen_400}
Top : the outgoing photon energy dependence of the
$\gamma p \rightarrow \gamma \pi^0 p$ 3-fold differential
{\it c.m.} cross section $d\sigma/dE_\gamma' d\Omega_\pi$,
divided by its value in the soft-photon limit,
as function of the outgoing photon energy
$E'^{cm}_\gamma$, at incident photon {\it lab} energy
$E_\gamma^{lab}=400$~MeV and
pion emission angle $\theta_\pi^{c.m.}=90^{o}$.
Predictions of the unitary model
for $\kappa_{\Delta^+}=0$ (dotted curve) and
$\kappa_{\Delta^+}=3$ (full curve).
Bottom : same for the photon asymmetry $\Sigma$.
The horizontally dashed curve at small values of $E'^{c.m.}_\gamma$ is
obtained using the low energy theorem, and corresponds to the photon
asymmetry for the $\gamma p \to \pi^0 p$ reaction.}
\end{figure}
%
\newline
\indent We next study how to extract new resonance information from the
deviations from the soft-photon limit in the $\gamma p \to \gamma \pi^0 p$
cross sections at the larger values of $E'_\gamma$. In
Fig.~\ref{fig:gap_gapiop_3fen_400} we show the sensitivity of the 3-fold
differential cross sections and the photon asymmetry, for linearly polarized 
incident photons, to the $\Delta^+(1232)$ MDM. We present both 
cross section and photon asymmetry 
at an incident photon energy of 400 MeV, for which our model yields a
good description of the $\gamma p \to \pi^0 p$ observables. This serves as a
reliable baseline to study the dependence of the $\gamma p \to \gamma \pi^0 p$
process on $\kappa_{\Delta^+}$ at the larger values of $E'_\gamma$. It is seen
from Fig.~\ref{fig:gap_gapiop_3fen_400} that an outgoing photon energy
$E'_\gamma$ of around 100 MeV is a good compromise to enhance the sensitivity
to $\kappa_{\Delta^+}$ while still staying in the region of validity of the
present calculation, which treats the radiation due to rescattering effects in
the soft-photon approximation.
%
\begin{figure}[h]
\includegraphics[width=0.5\columnwidth]{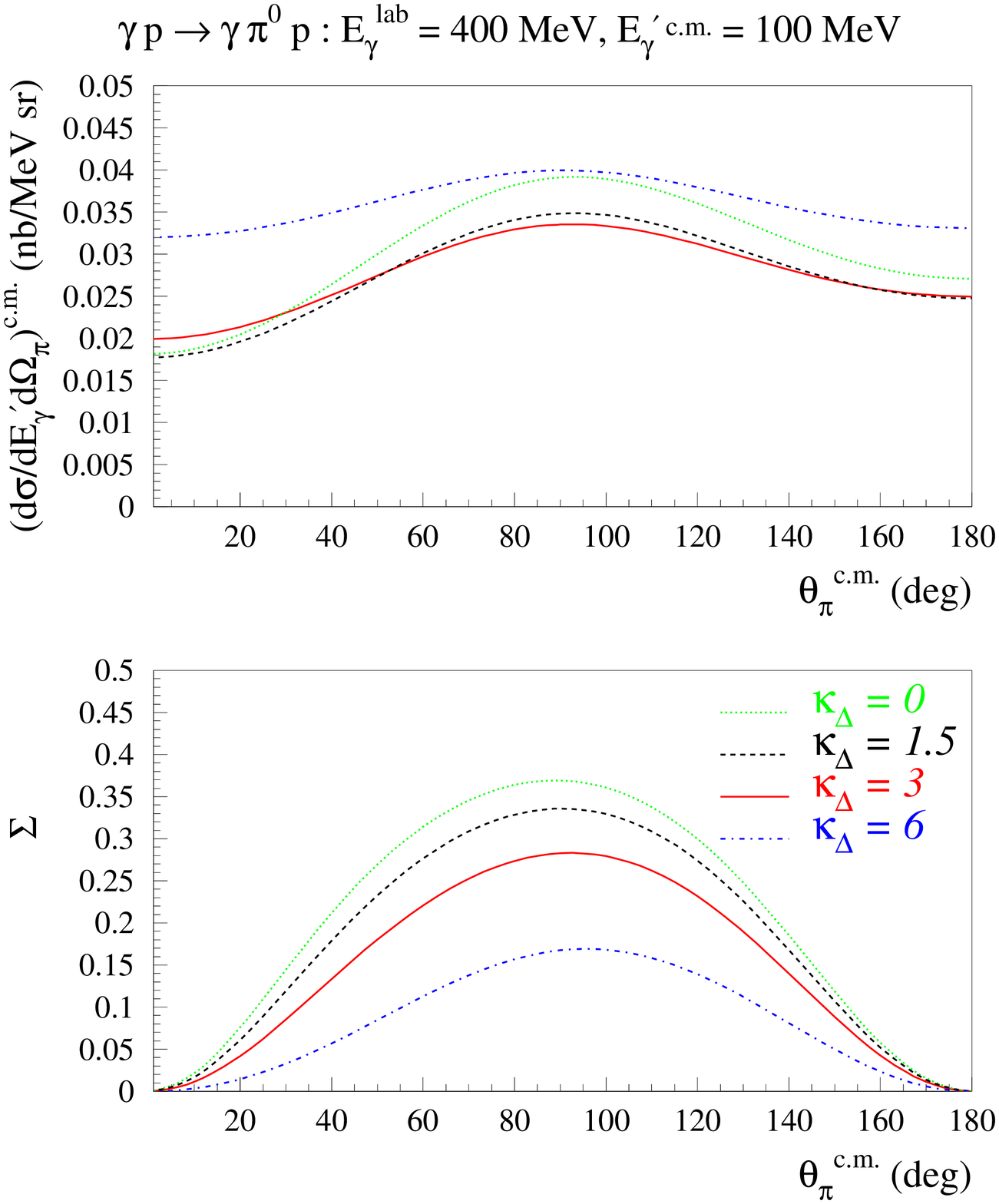}
\caption{
\label{fig:gap_gapiop_3fang_400}
Top : the angular distribution of the emitted pions for the $\gamma p\to \gamma
\pi^0 p$ 3-fold differential {\it c.m.} cross section $d\sigma/dE'_\gamma
d\Omega_\pi$ at incident photon lab energy $E_\gamma^{lab}=400$~MeV and fixed
outgoing photon energy $E'^{c.m.}_\gamma=100$~MeV. The sensitivity of the
unitary model to different values of $\kappa_{\Delta^+}$ is shown. Bottom :
same for the photon asymmetry $\Sigma$.  }
\end{figure}
%
\newline
\indent In Fig.~\ref{fig:gap_gapiop_3fang_400} we therefore investigate the
sensitivity of the pion angular distribution at $E^{lab}_\gamma = 400$~MeV and
$E'^{cm}_\gamma = 100$~MeV with regard to the value of $\kappa_{\Delta^+}$. The
upper part of Fig.~\ref{fig:gap_gapiop_3fang_400} shows a considerable change
in the angular distribution of the differential cross section when varying
$\kappa_{\Delta^+}$ between 0 and 6. However it is also obvious that extracting
a value of $\kappa_{\Delta^+}$ from a fit to the angular distribution would
require very accurate data over the whole angular range. The reason is that the
differential cross section first decreases when increasing $\kappa_{\Delta^+}$
from the value $\kappa_{\Delta^+} = 0$ , reaches a minimum around a value
$\kappa_{\Delta^+} = 3$, and increases subsequently when increasing
$\kappa_{\Delta^+}$ further. This behavior is due to interference and evidently
complicates an accurate extraction of $\kappa_{\Delta^+}$ from the differential
cross section. However, we found that the photon asymmetry, for linearly
polarized incident photons, decreases monotonically when increasing
$\kappa_{\Delta^+}$, as is displayed in the lower part of
Fig.~\ref{fig:gap_gapiop_3fang_400}. In particular, the photon asymmetry varies
between +0.35 and +0.15 when varying $\kappa_{\Delta^+}$ from 0 to 6.
%
\begin{figure}[h]
\includegraphics[width=0.65\columnwidth]{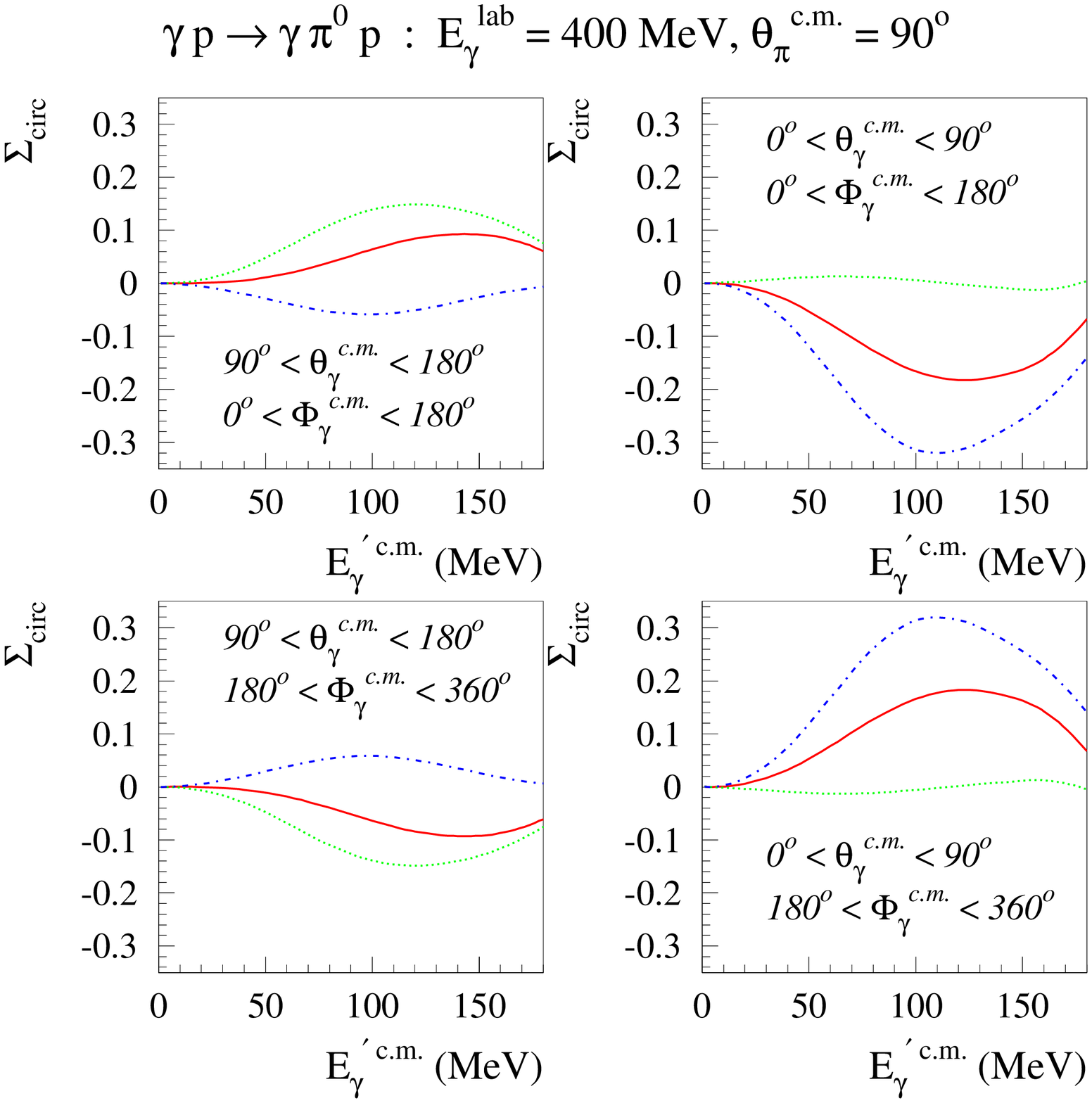}
\caption{
\label{fig:sigmacirc}
The outgoing photon energy dependence of the
$\gamma p \rightarrow \gamma \pi^0 p$
single spin asymmetry $\Sigma_{circ}$ for a circularly polarized
incident photon
at $\theta_\pi^{c.m.} = 90^o$ when integrating over the outgoing photon
angles as indicated on the figure.
Predictions of the unitary model are shown
for $\kappa_{\Delta^+}=0$ (dotted curves),
$\kappa_{\Delta^+}=3$ (solid curves),
and $\kappa_{\Delta^+}=6$ (dashed-dotted curves).
}
\end{figure}
%
\newline
\indent Besides the photon asymmetry, for linearly polarized incident photons,
we have also studied the single asymmetry for a circularly polarized incident
photon, which we denote as $\Sigma_{circ}$. For a two-body reaction, such as
$\gamma N \to \pi N$, parity conservation forces $\Sigma_{circ}$ to vanish
exactly because of the reflection symmetry with respect to the reaction plane.
For a three-body process, such as $\gamma N \to \gamma \pi N$, this reflection
symmetry is broken due to the emission of the second particle, and one can
define a single spin asymmetry for circularly polarized photons as~:
\begin{eqnarray}
\label{eq:sigmacirc}
\Sigma_{circ} \equiv \frac{d \sigma (\lambda = +1) - d \sigma (\lambda = -1)}
{d \sigma (\lambda = +1) + d \sigma (\lambda = -1) },
\end{eqnarray}
where $d \sigma$ in Eq.~(\ref{eq:sigmacirc}) stands for $\left( d \sigma / d
E'_\gamma d \Omega_\gamma d \Omega_\pi \right)^{c.m.}$, and $\lambda = \pm 1$
are the two circular polarization states of the incident photon. In
Fig.~\ref{fig:sigmacirc}, we show $\Sigma_{circ}$ for the $\gamma p \to \gamma
\pi^0 p$ reaction as function of the outgoing photon energy when the pion is
emitted at an angle $\theta_\pi^{c.m.} = 90^o$, which fixes the reaction plane.
One may then study the dependence of $\Sigma_{circ}$ when integrating $d
\sigma$ over the angles of the outgoing photon. In Fig.~\ref{fig:sigmacirc}, we
separated the phase space for the outgoing photon into 4 quadrants, as the
photon can be emitted in the forward or backward (with regard to the direction
of the incident photon) hemispheres, and either above or below (with regard to
the direction of $\vec k \times \vec q$) the reaction plane. One immediately
observes from Fig.~\ref{fig:sigmacirc} that the sign of $\Sigma_{circ}$ differs
for photons emitted above and below the reaction plane. As a result the
integral of $\Sigma_{circ}$ over the full solid angle of the final photon
vanishes because this way one effectively obtains the result of a two-body
reaction. A further interesting feature of Fig.~\ref{fig:sigmacirc} is that
$\Sigma_{circ}$ vanishes exactly in the soft-photon limit. This can also be
easily understood from the fact that in the soft-photon limit the LET relates
the $\gamma N \to \gamma \pi N$ process to the two-body reaction $\gamma N \to
\pi N$ for which $\Sigma_{circ}$ vanishes. Since the soft-photon emission from
the external charged particles does not contribute to $\Sigma_{circ}$, this
observable acts as a filter to enhance the $\Delta$-resonant process. Indeed,
one observes from Fig.~\ref{fig:sigmacirc} that in the forward and upper
quadrant (for $\Phi_\gamma > 0$) and at an energy $E'_\gamma = 100$~MeV,
$\Sigma_{circ}$ changes from 0 to -0.3 when $\kappa_{\Delta^+}$ is varied
between 0 and 6. Furthermore, in the backward and upper quadrant and at an
energy $E'_\gamma = 100$~MeV, $\Sigma_{circ}$ changes between +0.15 and -0.05
when varying $\kappa_{\Delta^+}$ between 0 and 6. Since circularly polarized
photons are readily available at MAMI, a measurement of $\Sigma_{circ}$ for the
$\gamma p \to \gamma \pi^0 p$ reaction in the $\Delta(1232)$ region provides a
unique opportunity to enhance the $\Delta$-resonant process and access
$\kappa_{\Delta^+}$.
%
\begin{figure}[h]
\includegraphics[width=0.5\columnwidth]{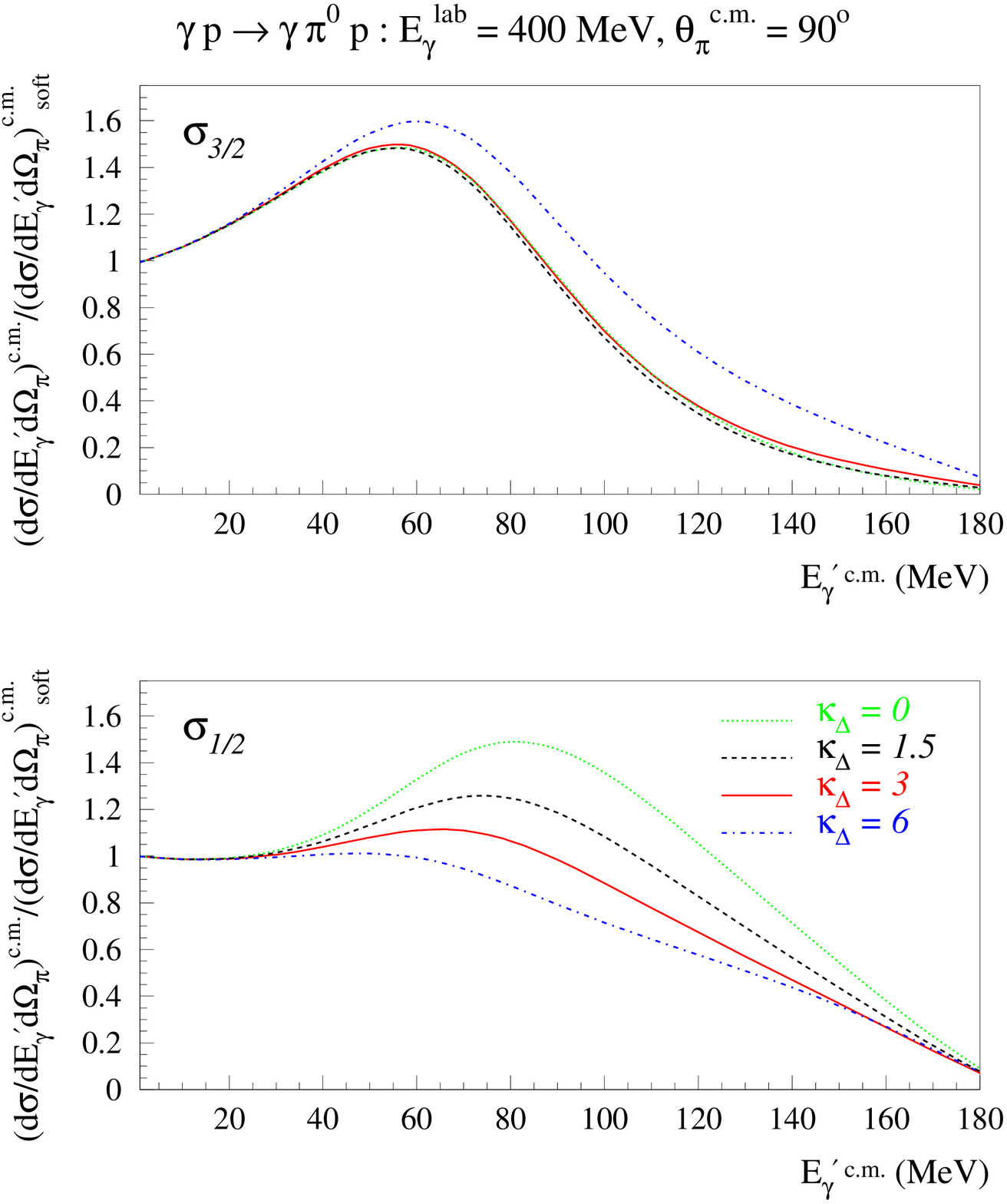}
\caption{
\label{fig:gap_gapiop_hel_en}
The helicity dependence of the $\gamma
p\rightarrow \gamma \pi^0 p$ {\it c.m.}
cross section $d\sigma/dE_\gamma' d\Omega_\pi$,
divided by its soft photon value, as function of the outgoing photon energy
$E'^{cm}_\gamma$, at incident photon {\it lab}
energy $E_\gamma^{lab}=400$~MeV and
pion emission angle $\theta_\pi^{cm}=90^{o}$.
Upper (lower) panel shows the cross sections for total helicity 3/2 (1/2)
respectively.
The curves correspond with the predictions of the
unitary model for different values of $\kappa_{\Delta^+}$ as
indicated on the figure.}
\end{figure}
%
%
\begin{figure}[h]
\includegraphics[width=0.5\columnwidth]{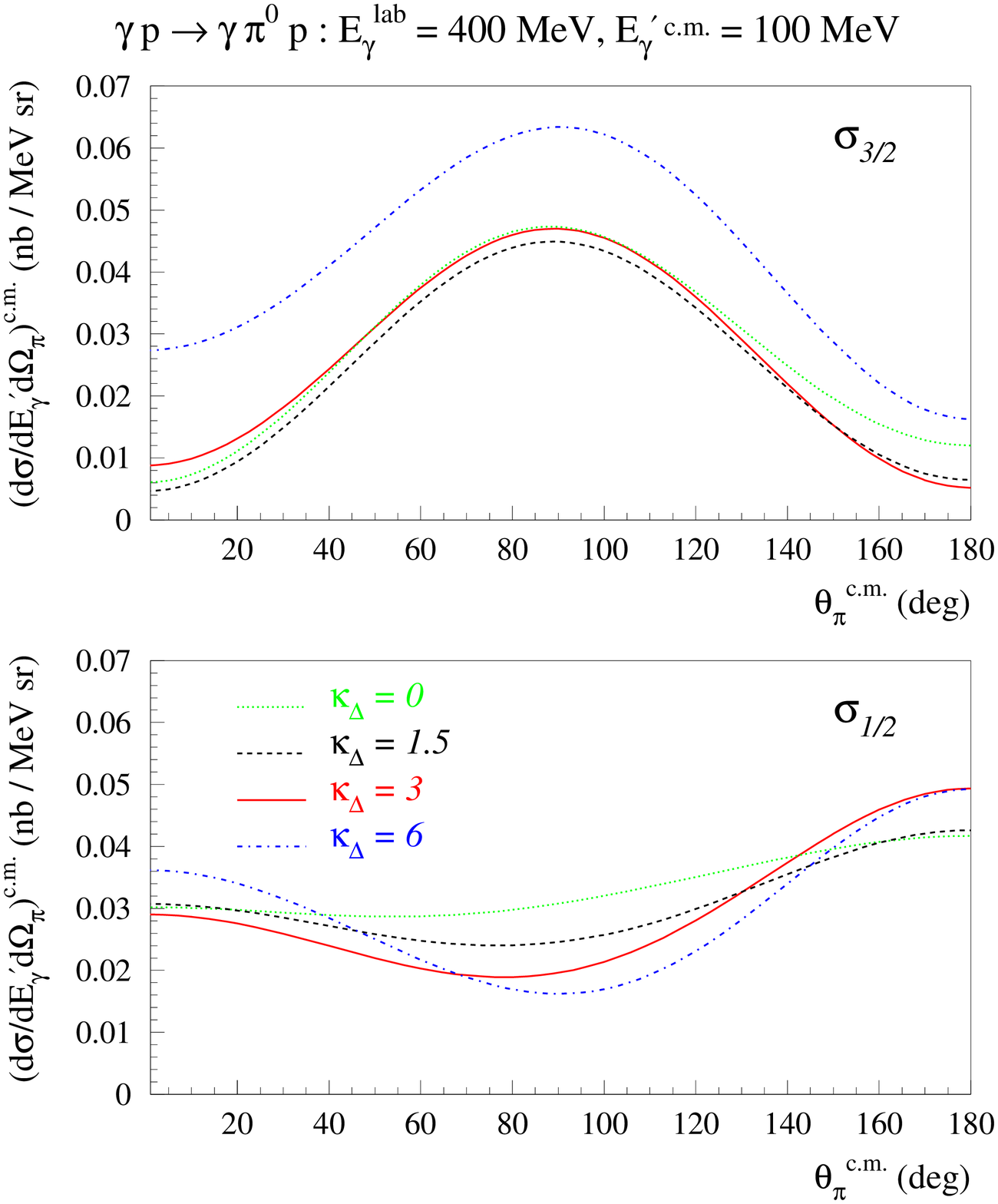}
\caption{
\label{fig:gap_gapiop_hel_ang}
The angular dependence of the emitted pions
for the $\gamma p \rightarrow \gamma \pi^0 p$ {\it c.m.}
helicity cross sections $d\sigma/dE_\gamma' d\Omega_\pi$,
at incident photon lab energy $E_\gamma^{lab}=400$~MeV and fixed outgoing
photon energy $E'^{c.m.}_\gamma=100$~MeV.
Upper (lower) panel shows the cross sections for total helicity 3/2 (1/2)
respectively.
The curves correspond with the predictions of the
unitary model for different values of $\kappa_{\Delta^+}$ as
indicated on the figure.}
\end{figure}
%
\newline
\indent We next investigate double spin observables where both the incident
photon and the target proton are polarized. In the following figures we display
the sensitivity of the total helicity cross sections for the $\gamma p \to
\gamma \pi^0 p$ reaction to the value of $\kappa_{\Delta^+}$.
Figure~\ref{fig:gap_gapiop_hel_en} shows the dependence of these cross sections
on the outgoing photon energy, and Fig.~\ref{fig:gap_gapiop_hel_ang} the pion
angular distributions. These helicity cross sections are accessible
experimentally by measuring the $\gamma p \to \gamma \pi^0 p$ reaction with a
circularly polarized photon beam and a longitudinally polarized proton target,
for the cases of parallel ($\sigma_{3/2}$) or anti-parallel ($\sigma_{1/2}$)
spins. It is seen from Fig.~\ref{fig:gap_gapiop_hel_en} that in the low energy
limit ($E_\gamma' \to 0$), one exactly recovers the helicity cross sections of
the $\gamma p \to \pi^0 p$ reaction, for which we obtained a good description (
see Fig.~\ref{fig:gap_piop_hel} ). At the higher values of $E_\gamma'$, one
notices an interference pattern in the $\sigma_{3/2}$ cross section that
strongly reduces the dependence on $\kappa_{\Delta^+}$ in the range from 0 to
3. The $\sigma_{1/2}$ cross section, on the other hand, decreases monotonically
with increasing $\kappa_{\Delta^+}$, thus indicating a very strong sensitivity
to the $\Delta^+$ MDM in the range 70~MeV$\le E_\gamma' \le 120$~MeV (see
Fig.~\ref{fig:gap_gapiop_hel_en}, bottom). Also the angular distribution of
$\sigma_{1/2}$ is very sensitive to the value of $\kappa_{\Delta^+}$, showing a
rather flat distribution for $\kappa_{\Delta^+}=0$ and a distinct minimum for
$\kappa_{\Delta^+}=6$ (see Fig.~\ref{fig:gap_gapiop_hel_ang}, bottom). A
measurement of these helicity cross sections will be feasible in the near
future at MAMI.
%
\begin{figure}[h]
\includegraphics[width=0.4\columnwidth]{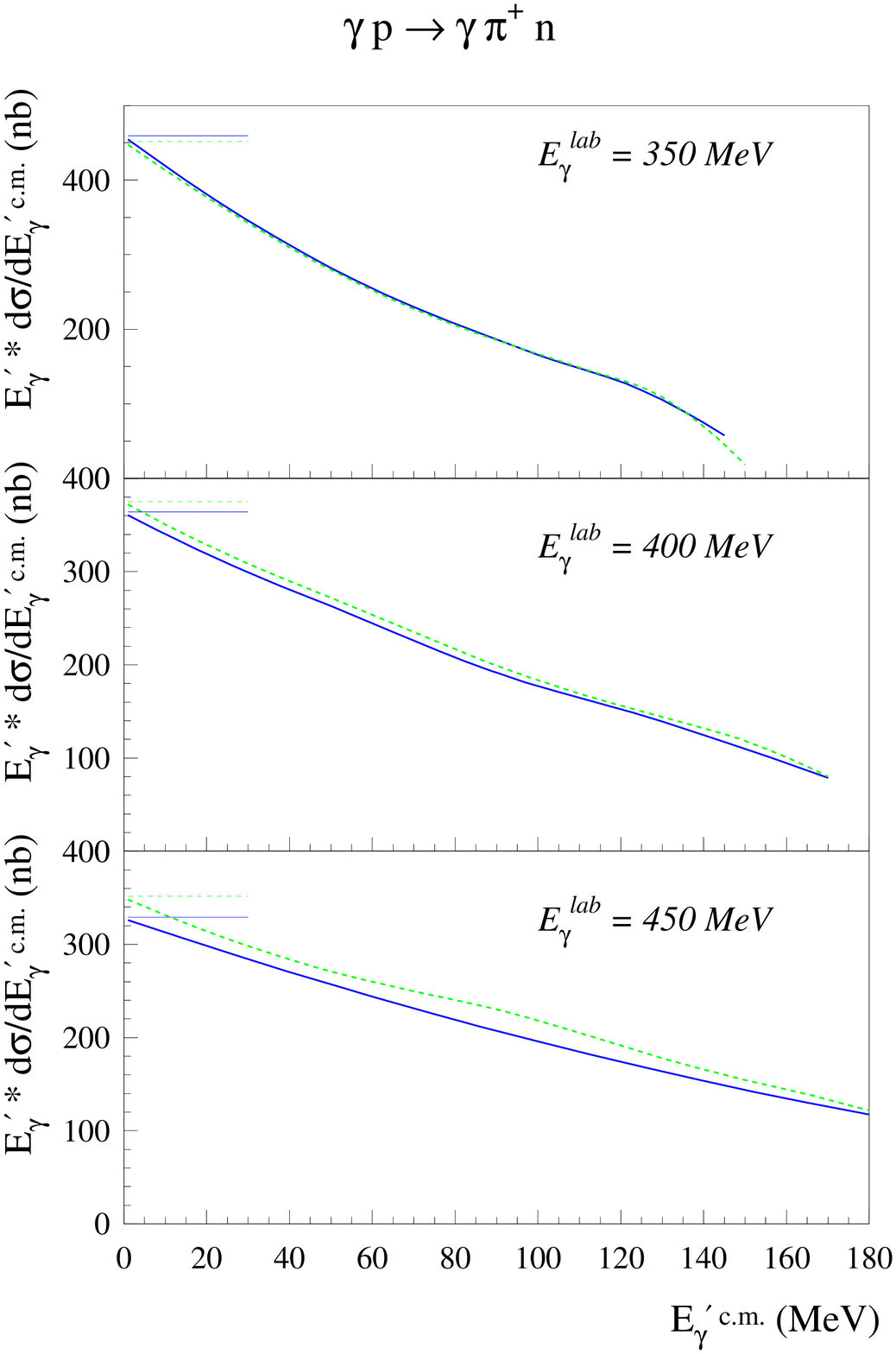}
\includegraphics[width=0.4\columnwidth]{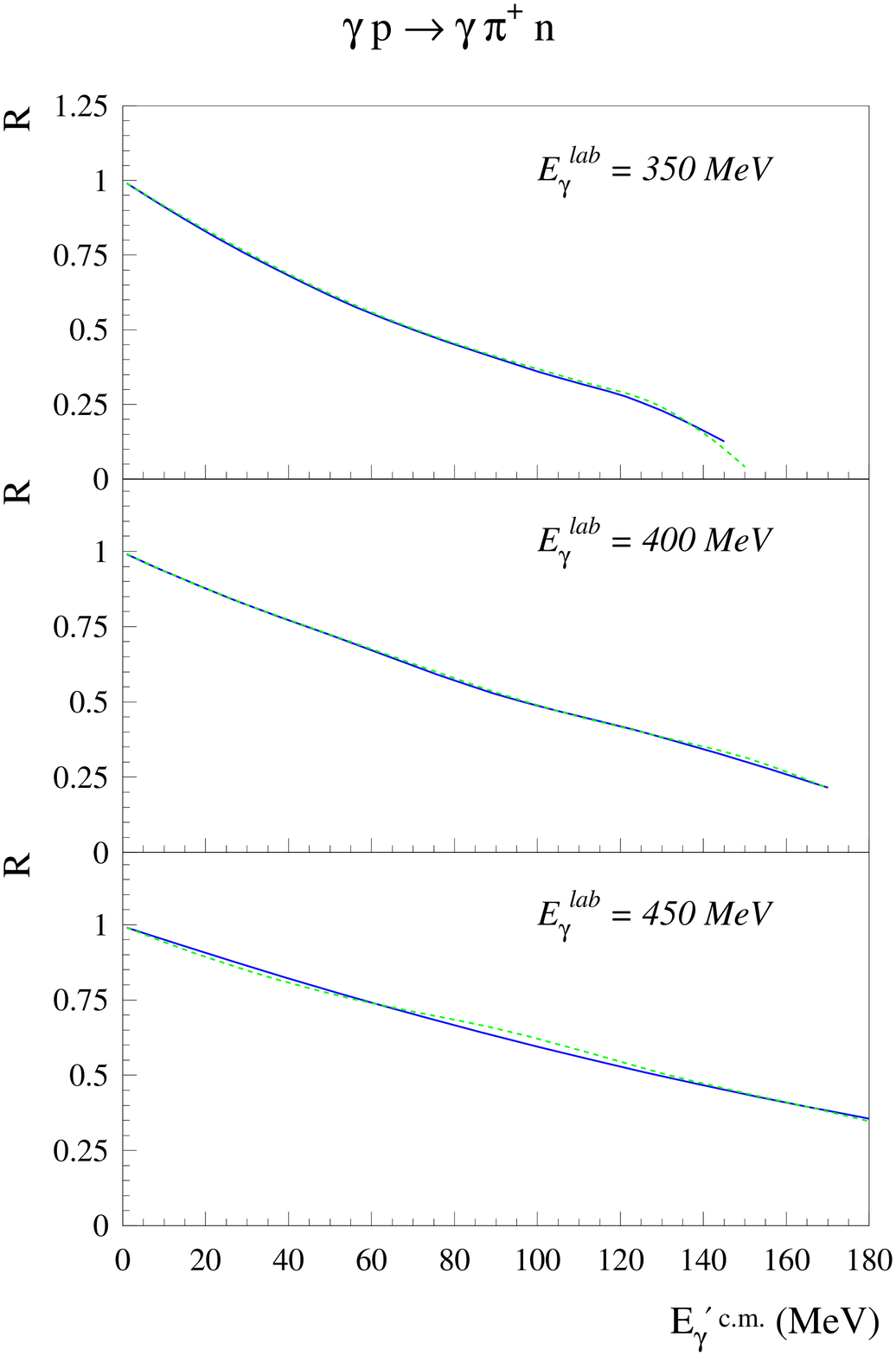}
\caption{\label{fig:gapipn_tcs}
Left panel : outgoing photon energy dependence of the
$\gamma p \to \gamma \pi^+ n$ cross section
$d\sigma/dE'^{c.m.}_\gamma$ multiplied by $E'^{c.m.}_\gamma$.
Right panel : ratio $R$ of the $\gamma p \rightarrow \gamma \pi^+ n$ process,
as defined in Eq.~(\ref{eq:Rpip}).
Notation as in Fig.~\ref{fig:gapiop_tcs}. }
\end{figure}
%
\newline
\indent Although the main focus of our work is a unitary model for the $\gamma
p \to \gamma \pi^0 p$ reaction in the $\Delta(1232)$ resonance region, we also
obtain, within the same framework, a description of the $\gamma p \to \gamma
\pi^+ n$ reaction. The tree level part will now contain additional terms where
the soft photon couples to the charged pion as shown in
Fig.~\ref{fig:gap_gapiN_diag}. The rescattering terms in the soft photon limit
for $\gamma p \to \gamma \pi^+ n$ are obtained by the replacement $p' \to q$ in
the second term of Eq.~(\ref{eq:tmatrix_gagapi}). As we have seen from the
results in Sec.~\ref{sec:pi}, the $\gamma p \to \pi^+ n$ reaction has a much
larger nonresonant contribution compared with the $\gamma p \to \pi^0 p$
reaction. For extracting information on the $\Delta^+$ MDM, the $\gamma p \to
\gamma \pi^0 p$ reaction is therefore clearly the favorite channel. However,
the present experiments of Ref.~\cite{CB} will simultaneously measure both
$\gamma p \to \gamma \pi^0 p$ and $\gamma p \to \gamma \pi^+ n$ reactions.
Therefore, the $\gamma p \to \gamma \pi^+ n$ data may provide a useful
additional cross-check for our theoretical description. In
Fig.~\ref{fig:gapipn_tcs}, we show the outgoing photon energy dependence of the
cross section $d\sigma/dE'_\gamma$ for the $\gamma p \to \gamma \pi^+ n$
reaction, integrated over the photon and pion angles for three incoming photon
energies through the $\Delta(1232)$ region. By comparing the left panels of
Figs.~\ref{fig:gapiop_tcs} and \ref{fig:gapipn_tcs}, we observe that at small
outgoing photon energies, the $\gamma p \to \gamma \pi^+ n$ cross sections are
about a factor 10 larger than the corresponding $\gamma p \to \gamma \pi^0 p$
cross sections. This is readily understood by the fact that in the soft-photon
limit there is a large contribution due to radiation from the charged pion for
the $\gamma p \to \gamma \pi^+ n$ process. On the other hand, for the $\gamma p
\to \gamma \pi^0 p$ process only bremsstrahlung contributions arise from the
emission of soft photons from the (much heavier) protons.
\newline
\indent
Similar as in Fig.~\ref{fig:gapiop_tcs}, we can also construct the
ratio R between the $\gamma p \to \gamma \pi^+ n$ process and its soft-photon
limit as given by the LET, which is derived in Appendix~\ref{sec:LET2}, see
Eq.~(\ref{eq:Rpip}). In contrast to the $\gamma p \to \gamma \pi^0 p$ process,
where this ratio shows clear resonance structure when increasing the final
photon energy, the corresponding ratio for the $\gamma p \to \gamma \pi^+ n$
process drops monotonously with increasing $E'_\gamma$.
\newline
\indent
In Fig.~\ref{fig:gap_gapipn_hel_ang}, we show the pion angular
dependence of the 3-fold differential cross section and photon asymmetry for
the $\gamma p \to \gamma \pi^+ n$ reaction at $E_\gamma = 400$~MeV and
$E'_\gamma = 100$~MeV. It is seen from Fig.~\ref{fig:gap_gapipn_hel_ang} that
for an energy above the $\Delta(1232)$ resonance position, the $\gamma p \to
\gamma \pi^+ n$ differential cross section exhibits a forward peaking analogous
to the $\gamma p \to \pi^+ n$ process. Furthermore, it is seen that both the
differential cross section and the photon asymmetry for the $\gamma p \to
\gamma \pi^+ n$ reaction only display a rather modest change when varying
$\kappa_{\Delta^+}$ between 0 and 6. Therefore, the measurement of the $\gamma
p \to \gamma \pi^+ n$ process can put stringent constraints on our theoretical
description of the non-resonant contributions, making it a useful tool to
minimize model dependencies when extracting information on the $\Delta^+$ MDM
from the $\gamma p \to \gamma \pi^0 p$ process.
%
\begin{figure}[h]
\includegraphics[width=0.45\columnwidth]{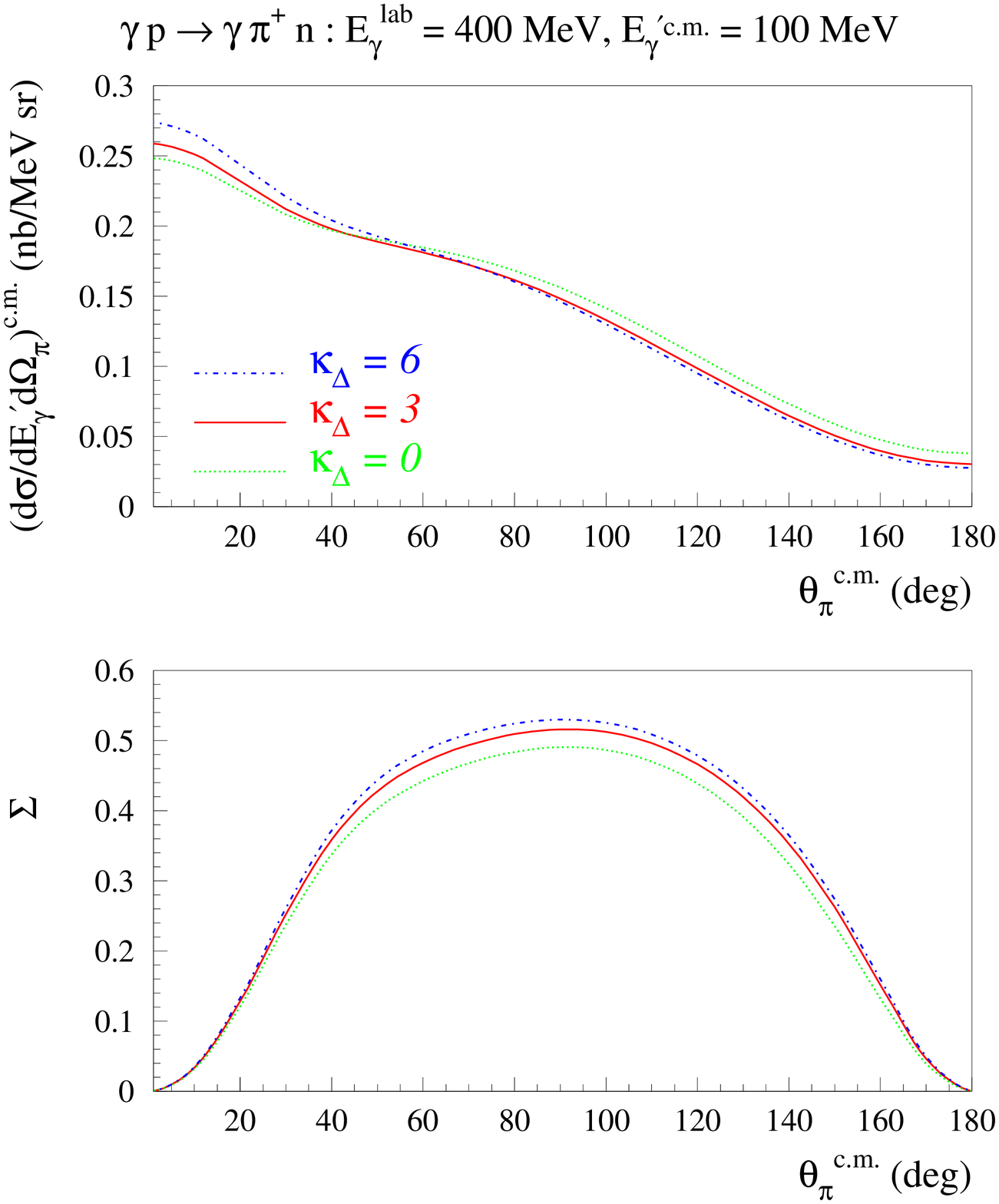}
\caption{
\label{fig:gap_gapipn_hel_ang}
Top : the angular distribution of the emitted pions for the $\gamma p\to \gamma
\pi^+ p$ 3-fold differential {\it c.m.} cross section $d\sigma/dE'_\gamma
d\Omega_\pi$ at incident photon lab energy $E_\gamma^{lab}=400$~MeV and fixed
outgoing photon energy $E'^{c.m.}_\gamma=100$~MeV. The sensitivity of the
unitary model to different values of $\kappa_{\Delta^+}$ is shown. Bottom :
same for the photon asymmetry $\Sigma$. }
\end{figure}
%

\section{Conclusions and outlook}
\label{sec:concl}

In this work, we provided a unitary model for the $\gamma p \to \gamma \pi N$
reaction in the $\Delta(1232)$ region. Our starting point is a
unitary model for the $\gamma p \to \pi N$ ($\pi N = \pi^0 p, \pi^+ n$)
reaction based on a transition potential consisting of
Born diagrams, vector meson exchanges and $\Delta$-resonant process.
The rescattering effects are included in an on-shell (K-matrix) approximation.
Besides the vector meson coupling constants, the only free parameters in this
model are the $\gamma N \Delta$ electric and magnetic couplings.
With this model we find a very reasonable description of both total
unpolarized and helicity cross sections
as well as differential cross sections and photon asymmetries
for both the $\gamma p \to \pi^0 p$ and $\gamma p \to \pi^+ n$
processes through the $\Delta(1232)$ resonance region.
\newline
\indent
The model for the $\gamma p \to \pi N$ processes was then
extended  to describe the $\gamma p \to \gamma \pi N$ reactions.
Our model for these reactions is gauge invariant
with respect to both initial and final photons.
In particular, it is constructed such that in the limit of
small outgoing photon energy, it exactly reproduces the low energy theorem
which relates the $\gamma p \to \gamma \pi N$ and $\gamma p \to \pi N$
processes.
For the $\gamma p \to \gamma \pi N$ reactions,
the tree level terms include Born diagrams, vector meson exchanges, the
$\pi^0 \to \gamma \gamma$ anomaly contribution and the $\Delta(1232)$
contribution.
The rescattering effects are calculated in the soft-photon approximation using
the previously described model for $\gamma p \to \pi N$.
In this framework, the only new parameter entering in the description of the
$\gamma p \to \gamma \pi N$ process is the $\Delta^+(1232)$
magnetic dipole moment (MDM).
\newline
\indent Using this unitary model, we investigated several $\gamma p \to \gamma
\pi N$ observables. We found good agreement for the existing experimental data
of the $\gamma p \to \gamma \pi^0 p$ reaction. In particular, the rescattering
effects were found to slightly reduce the cross sections at the larger outgoing
photon energies, which improved the agreement with the data. We then studied
the sensitivity of the $\gamma p \to \gamma \pi^0 p$ observables to the
$\Delta^+$ MDM. The unpolarized differential cross section displays an
interference structure which reduces the sensitivity to values of
$\kappa_{\Delta^+}$ in the range between 0 and 3. However, the photon asymmetry
for linearly polarized incident photons is strongly dependent on
$\kappa_{\Delta^+}$, and changes between 0.35 and 0.15 when $\kappa_{\Delta^+}$
is varied between 0 and 6. The dedicated measurements of the photon asymmetry
that are currently underway at MAMI are therefore highly promising for a more
quantitative extraction of the $\Delta^+$ MDM.
\newline
\indent The single asymmetry for circularly polarized incident photons provides
a new observable for a three-body reaction such as $\gamma p \to \gamma \pi N$
if the photon is emitted out of the plane defined by the incident photon and
the final pion. This asymmetry has the interesting feature that it vanishes
exactly in the soft-photon limit, where the $\gamma p \to \gamma \pi^0 p$
process effectively reduces to a two-body reaction, for which the asymmetry
vanishes. Since the pure bremsstrahlung contribution due to soft-photon
emission from the external charged particles does not contribute, the single
asymmetry for circularly polarized incident photons therefore acts as a filter
to enhance the $\Delta$-resonant process, and indeed our results display a
strong sensitivity to the $\Delta^+$ MDM.
\newline
\indent Yet another sensitive observable with regard to the $\Delta^+$ MDM is
the helicity cross section $\sigma_{1/2}$. In particular, this differential
cross section changes by about a factor of 2 when $\kappa_{\Delta^+}$ is varied
between 0 and 6.
\newline
\indent
Besides a prediction for the $\gamma p \to \gamma \pi^0 p$ observables,
our unitary model also provides
a description of the $\gamma p \to \gamma \pi^+ n$ reaction.
However, the $\gamma p \to \gamma \pi^+ n$ process is
dominated by non-resonant processes and bremsstrahlung contributions
originating from radiation off the charged pion line.
Therefore, the measurement of the $\gamma p \to \gamma \pi^+ n$ process
will put stringent constraints on our theoretical description of
the non-resonant contributions, and can be useful
in minimizing model dependencies when extracting information on the
$\Delta^+$ MDM from the $\gamma p \to \gamma \pi^0 p$ process.
\newline
\indent
To improve on the accuracy in the extraction of the
$\Delta^+$ MDM, the present framework may be extended
to include the rescattering corrections
at finite final photon energies. At such energies
the final photon can not only be emitted from the external charged lines as
in Figs.~\ref{fig:rescatter} (b) and (c),
but also from an intermediate line.
A particularly important contribution in the $\Delta$-region is expected
to come from the $\Delta \to \pi N \to \Delta$ self-energy contribution
where the photon is emitted from the light pion in the loop.
The Ward identity then requires the energy dependence of this
vertex correction be consistent with the energy dependence of the
$\Delta$ self energy. Such a calculation, which
will provide an imaginary part to the $\gamma \Delta \Delta$ vertex,
is beyond the scope of the present work.
However, it can be worked out in the future \cite{Pas}
as an input for the present framework
in order to extend its range of applicability.
\newline
\indent
The upcoming dedicated measurements of the
$\gamma p \to \gamma \pi N$ reaction will certainly trigger
new theoretical efforts with the aim to further minimize
the model dependencies in the extraction of the $\Delta^+$ MDM.

\begin{acknowledgments}

The authors would like to thank R. Beck, M. Kotulla,
and V. Pascalutsa for helpful discussions.
W.-T.~C. is grateful to the Universit\"at Mainz for the hospitality
extended to him during his visits.
This work was supported in parts by the
National Science Council of ROC under grant
No.~NSC92-2112-M002-013 and
NSC92-2112-M-001-058, the Deutsche Forschungsgemeinschaft (SFB 443),
a joint project NSC/DFG TAI-113/10/0,
and the US Department of Energy
under contracts DE-FG02-04ER41302 and DE-AC05-84ER40150.

\end{acknowledgments}

\appendix

\section{Derivation of the low energy theorems}
\label{app:soft}%

\subsection{Low energy theorem relating the $\bm{\gamma p \to \gamma
\pi^0 p}$ and $\bm{\gamma p \to \pi^0 p}$ processes}
\label{sec:LET1}
In the soft-photon limit the $\gamma p \rightarrow \gamma \pi^0 p$ reaction is
exactly described by the bremsstrahlung process from the initial and final
protons. This yields the following five-fold differential {\it c.m.}
cross section in the limit $k' \to 0$:
\begin{equation} \label{eq:softa1}
\left(  \frac{d\sigma}{dE'_\gamma d\Omega_\pi
                 d\Omega_\gamma} \right)^{c.m.}
  \;\longrightarrow\;
  \frac{e^2}{16 \pi^3} \, E'_\gamma \,
  \sum_{\lambda'} \left| \,
  \frac{p' \cdot \varepsilon(k', \lambda')}{p' \cdot k'}
  \;-\;
  \frac{p \cdot \varepsilon(k', \lambda')}{p \cdot k'}
  \, \right|^2
  \left(\frac{d\sigma}{d\Omega_\pi}\right)^{c.m.} \,,
\end{equation}
where $(d\sigma / d\Omega_\pi)^{c.m.}$ is the {\it c.m.} cross section for
the $\gamma p \rightarrow \pi^0 p$ process, $\lambda' = \pm 1$ the photon
polarization, and $\varepsilon$ its polarization vector. We calculate the
{\it rhs}
of Eq.~(\ref{eq:softa1}) by performing the sum over the photon polarizations
and integrating over the photon angles. This gives the result
\begin{equation}
\label{eq:softa2}
\left(  \frac{d\sigma}{dE'_\gamma d\Omega_\pi } \right)^{c.m.}
 \; \longrightarrow \;
  \frac{e^2}{16 \pi^3} \, E'_\gamma \; \mathcal{I}
  \left(\frac{d\sigma}{d\Omega_\pi}\right)^{c.m.} \,,
\end{equation}
where we introduced the photon angular integral $\mathcal{I}$ as
\begin{equation} \label{eq:softa3}
  \mathcal{I} \,\equiv \, \int d \Omega^{c.m.}_\gamma
  \left[ \, \frac{2 \, p \cdot p'}{(p \cdot k') \, (p' \cdot k')}
  \,-\, \frac{M_N^2}{(p \cdot k')^2} \,-\, \frac{M_N^2}{(p' \cdot k')^2} \,
  \right] \,.
\end{equation}
In Eq.~(\ref{eq:softa3}),
the second and third terms arise from the contribution of
brems\-strah\-lung from the initial and final proton, respectively, whereas the
first term stems from the interference between the bremsstrahlung
amplitudes from the initial and final protons.
\newline
\indent
We next work out the photon angular integral of Eq.~(\ref{eq:softa3}). It is
convenient to introduce the initial and final nucleon velocities $\bm{\beta_N}
\equiv \bm{p} / E_N$ and $\bm{\beta'_N} \equiv \bm{p'} / E'_N$, and a
Feynman parametrization of the first term of Eq.~(\ref{eq:softa3}), which
brings the propagators to the same denominator. The result is
\begin{eqnarray} \label{eq:softa4}
  \mathcal{I} &=& \frac{2 \pi}{(E'_\gamma)^2 }
  \left\{ (1 - \bm{\beta_N} \cdot \bm{\beta'_N}) \, \int_{-1}^{+1}
  dy \, \int_{-1}^{+1} dx \, \frac{1}{(1 - \beta_y \, x)^2} \right.
  \\ \nonumber
& & \ \left.
  \,-\, (1 - \beta_N^2) \int_{-1}^{+1} dx \, \frac{1}{(1 - \beta_N \, x)^2}
  \,-\, (1 - \beta'^2_N) \int_{-1}^{+1} dx \,
  \frac{1}{(1 - \beta'_N \, x)^2} \right\} \,,
\end{eqnarray}
where $\beta_N$, $\beta'_N$, and $\beta_y$ are the magnitudes of
$\bm{\beta_N}$, $\bm{\beta'_N}$, and $\bm{\beta_y}$, which are related by
\begin{equation} \label{eq:softa5}
  \bm{\beta_y} \,\equiv\, \bm{\beta_N} \, \frac{1}{2} (1 + y) \;+\;
  \bm{\beta'_N} \, \frac{1}{2} (1 - y) \,.
\end{equation}
By use of the identity
\begin{equation} \label{eq:softa6}
  \int_{-1}^{+1} dx \, \frac{1}{(1 - \beta \, x)^2}
  \,=\, \frac{2}{1 - \beta^2} \,,
\end{equation}
Eq.~(\ref{eq:softa4}) can be cast into the form
\begin{equation} \label{eq:softa7}
  \mathcal{I} = \frac{2 \pi}{(E'_\gamma)^2 }
  \left\{ -4 \,+\, 2 \, (1 - \bm{\beta_N} \cdot \bm{\beta'_N}) \,
  \int_{-1}^{+1} dy \, \frac{1}{(1 - \beta_y^2)} \right\} \,.
\end{equation}
We next define the variable
\begin{equation}
v \equiv \sqrt{1 + \frac{4 M_N ^2}{-t } } ,
\end{equation}
with $t = (p' - p)^2$. This allows us to derive the relations
\begin{equation} \label{eq:softa8}
  (1 - \bm{\beta_N} \cdot \bm{\beta'_N}) =
  (1 - \beta_N^2)^{1/2} \, (1 - {\beta'_N}^2)^{1/2}
  \left( \frac{v^2 + 1}{v^2 - 1} \right)
\end{equation}
and
\begin{equation} \label{eq:softa9}
  \int_{-1}^{+1} dy \, \frac{1}{(1 - \beta_y^2 )} =
  \frac{1}{(1 - \beta_N^2)^{1/2} \,(1 - \beta'^2_N)^{1/2}} \,\cdot \,
  \left( \frac{v^2 - 1}{2 v} \right) \, \ln\!\left( \frac{v +
1}{v - 1} \right)^2 \,.
\end{equation}
Combining Eqs.~(\ref{eq:softa6}-\ref{eq:softa9}), we finally obtain
\begin{equation} \label{eq:softa10}
  \mathcal{I} = \frac{8 \pi}{(E'_\gamma)^2 }
  \left[ -1 \,+\, \left( \frac{v^2 + 1}{2 v} \right) \, \cdot \,
  \ln \left( \frac{v + 1}{v - 1} \right) \right] \,.
\end{equation}
If we insert this expression into the soft-photon limit for the cross section
of Eq.~(\ref{eq:softa2}), Eq.~(\ref{eq:softint2}) follows immediately.

\subsection{Low energy theorem relating the $\bm{\gamma p \to \gamma
\pi^+ n}$ and $\bm{\gamma p \to \pi^+ n}$ processes}
\label{sec:LET2}
The low-energy limit of $\pi^+$ production may be derived from the previous
results by a simple consideration. Whereas in the case of $\pi^0$ production
all final-state radiation comes from the protons,
in the case of $\pi^+$ production the charged pion
in the final state will radiate. Hence we obtain the relevant photon angular
integral from $\mathcal{I}$ of Eq.~(\ref{eq:softa3}) by replacing $p'\to q$
and, in the third term, $M_N^2\to m_{\pi}^2$. The result is
\begin{equation}
\label{eq:softb2}
\left(  \frac{d\sigma}{dE'_\gamma d\Omega_\pi } \right)^{c.m.}
  \longrightarrow
  \frac{e^2}{16 \pi^3} \, E'_\gamma \; \tilde{\mathcal{I}}
  \left(\frac{d\sigma}{d\Omega_\pi}\right)^{\!\mathrm{c.m.}} \,,
\end{equation}
with
\begin{equation} \label{eq:softb3}
  \tilde{\mathcal{I}} \,\equiv \, \int d \Omega^{\mathrm{c.m.}}_\gamma
  \left[ \, \frac{2 \, p \cdot q}{(p \cdot k') \, (q \cdot k')}
  \,-\, \frac{M_N^2}{(p \cdot k')^2} \,-\, \frac{m_\pi^2}{(q \cdot k')^2} \,
  \right] \,,
\end{equation}
where the second and third terms arise from the contribution of
brems\-strah\-lung from the initial proton and final pion alone, whereas the
first term stems from the interference between the
bremsstrahlung amplitudes from the initial proton and the final pion.

We next work out the photon angular integral of Eq.~(\ref{eq:softb3}) along the
same lines as in the case of a neutral pion. We introduce the initial nucleon
velocity $\bm{\beta_N} \equiv \bm{p} / E_N$ and the final pion velocity
$\bm{\beta_\pi} \equiv \bm{q} / \omega_\pi$, define an appropriate variable
$\tilde{v}$ as
\begin{equation}
\label{eq:v2}
\tilde{v} \equiv \sqrt{ 1 + \frac{4 M_N m_\pi}{(M_N-m_\pi)^2 - u} } \,,
\end{equation}
where $u \equiv (p - q)^2$.
The bremsstrahlung integral of Eq.~(\ref{eq:softb3}) can then be worked out
analogously as in Eq.~(\ref{eq:softa10}) and yields~:
\begin{equation} \label{eq:softb10}
  \tilde{\mathcal{I}} = \frac{8 \pi}{(E'_\gamma)^2 }
  \left[ -1 \,+\, \left( \frac{\tilde{v}^2 + 1}{2 \tilde{v}} \right) \,
  \ln \left( \frac{\tilde{v} + 1}{\tilde{v} - 1} \right) \right] \,.
\end{equation}
We note as an additional check that the new variable $\tilde{v}$ turns into the
previous variable $v$ if we replace $m_{\pi}\to M_N$ and $q\to p'$.
\newline
\indent
Finally we can also introduce the ratio $R$ between the
$\gamma p \to \gamma \pi^+ p$ process in the limit of a soft outgoing photon
and the $\gamma p \to \pi^+ n$ process as~:
\begin{eqnarray}
\label{eq:Rpip}
  R \,\equiv \, \frac{1}{\sigma_{\pi^+}} \cdot
  E'_\gamma
  \frac{d\sigma}{dE'_\gamma}\rightarrow 1\quad
  {\mbox{for}}\ E'_\gamma \to 0\, ,
\end{eqnarray}
with $\sigma_{\pi^+}$ defined as~:
\begin{equation}
\label{eq:spip}
  \sigma_{\pi^+}
  \equiv
  \frac{e^2}{2 \pi^2} \int d\Omega^{c.m.}_\pi\, W(\tilde v)
  \left(\frac{d\sigma}{d\Omega_\pi}\right)^{c.m.}
  \, .
\end{equation}
In Eq.~(\ref{eq:spip}), $W(\tilde v)$ is obtained from Eq.~(\ref{eq:wsoft})
by making the replacement $v \to \tilde v$.

\end{document}